\documentclass{bmcart}

\usepackage[utf8]{inputenc} %unicode support
\usepackage{amsmath}
\usepackage{amssymb}
\usepackage{url}
\usepackage{graphicx}

\startlocaldefs
\endlocaldefs

\begin{document}

%%% Start of article front matter
\begin{frontmatter}

\begin{fmbox}
\dochead{Research}


\title{Can co-location be used as a proxy for face-to-face contacts?}

%%%%%%%%%%%%%%%%%%%%%%%%%%%%%%%%%%%%%%%%%%%%%%
%%                                          %%
%% Enter the authors here                   %%
%%                                          %%
%% Specify information, if available,       %%
%% in the form:                             %%
%%   <key>={<id1>,<id2>}                    %%
%%   <key>=                                 %%
%% Comment or delete the keys which are     %%
%% not used. Repeat \author command as much %%
%% as required.                             %%
%%                                          %%
%%%%%%%%%%%%%%%%%%%%%%%%%%%%%%%%%%%%%%%%%%%%%%

\author[
   addressref={aff1,aff2},              % id's of addresses, e.g. {aff1,aff2}
   corref={aff2},                       % id of corresponding address, if any
%   noteref={},                          % id's of article notes, if any
   email={mathieu.genois@gesis.org}     % email address
]{\inits{M}\fnm{Mathieu} \snm{G\'enois}}
\author[
   addressref={aff1,aff3},
   email={alain.barrat@cpt.unv-mrs.fr}
]{\inits{A}\fnm{Alain} \snm{Barrat}}

%%%%%%%%%%%%%%%%%%%%%%%%%%%%%%%%%%%%%%%%%%%%%%
%%                                          %%
%% Enter the authors' addresses here        %%
%%                                          %%
%% Repeat \address commands as much as      %%
%% required.                                %%
%%                                          %%
%%%%%%%%%%%%%%%%%%%%%%%%%%%%%%%%%%%%%%%%%%%%%%

\address[id=aff1]{%                           % unique id
  \orgname{Aix Marseille Univ, Universit\'e de Toulon, CNRS, CPT}, % university, etc
  %\street{Waterloo Road},                     %
 % \postcode{13288}                            % post or zip code
  \city{Marseille},                           % city
  \cny{France}                                % country
}
\address[id=aff2]{%
  \orgname{GESIS, Leibniz Institute for the Social Sciences},
  \street{Unter Sachsenhausen 6-8},
  \postcode{50667}
  \city{K\"oln},
  \cny{Germany}
}
\address[id=aff3]{%
  \orgname{Data Science Laboratory, ISI Foundation},
  %\street{},
  %\postcode{50667}
  \city{Torino},
  \cny{Italy}
}

%%%%%%%%%%%%%%%%%%%%%%%%%%%%%%%%%%%%%%%%%%%%%%
%%                                          %%
%% Enter short notes here                   %%
%%                                          %%
%% Short notes will be after addresses      %%
%% on first page.                           %%
%%                                          %%
%%%%%%%%%%%%%%%%%%%%%%%%%%%%%%%%%%%%%%%%%%%%%%

\begin{artnotes}
%\note{Sample of title note}     % note to the article
%\note[id=n1]{Equal contributor} % note, connected to author
\end{artnotes}

\end{fmbox}% comment this for two column layout

%%%%%%%%%%%%%%%%%%%%%%%%%%%%%%%%%%%%%%%%%%%%%%
%%                                          %%
%% The Abstract begins here                 %%
%%                                          %%
%% Please refer to the Instructions for     %%
%% authors on http://www.biomedcentral.com  %%
%% and include the section headings         %%
%% accordingly for your article type.       %%
%%                                          %%
%%%%%%%%%%%%%%%%%%%%%%%%%%%%%%%%%%%%%%%%%%%%%%

\begin{abstractbox}

\begin{abstract} % abstract
%\parttitle{First part title} %if any
%Text for this section.

%\parttitle{Second part title} %if any
%Text for this section.
Technological advances have led to a strong increase in the number of data collection efforts aimed at measuring co-presence of individuals at different spatial resolutions. It is however unclear how much co-presence data can inform us on actual face-to-face contacts, of particular interest to study the structure of a population in social groups or for use in data-driven models of information or epidemic spreading processes. Here, we address this issue by leveraging data sets containing high resolution face-to-face contacts as well as a coarser spatial localisation of individuals, both temporally resolved, in various contexts. The co-presence and the face-to-face contact temporal networks share a number of structural and statistical features, but the former is (by definition) much denser than the latter. We thus consider several down-sampling methods that generate surrogate contact networks from the co-presence signal and compare them with the real face-to-face data. We show that these surrogate networks reproduce some features of the real data but are only partially able to identify the most central nodes of the face-to-face network. We then address the issue of using such down-sampled co-presence data in data-driven simulations of epidemic processes, and in identifying efficient containment strategies. We show that the performance of the various sampling methods strongly varies depending on context. We discuss the consequences of our results with respect to data collection strategies and methodologies.
\end{abstract}

%%%%%%%%%%%%%%%%%%%%%%%%%%%%%%%%%%%%%%%%%%%%%%
%%                                          %%
%% The keywords begin here                  %%
%%                                          %%
%% Put each keyword in separate \kwd{}.     %%
%%                                          %%
%%%%%%%%%%%%%%%%%%%%%%%%%%%%%%%%%%%%%%%%%%%%%%

\begin{keyword}
\kwd{face-to-face contacts}
\kwd{co-presence}
\kwd{digital epidemiology}
\kwd{complex networks}
\end{keyword}

% MSC classifications codes, if any
%\begin{keyword}[class=AMS]
%\kwd[Primary ]{}
%\kwd{}
%\kwd[; secondary ]{}
%\end{keyword}

\end{abstractbox}
%
%\end{fmbox}% uncomment this for twcolumn layout

\end{frontmatter}

%%%%%%%%%%%%%%%%%%%%%%%%%%%%%%%%%%%%%%%%%%%%%%
%%                                          %%
%% The Main Body begins here                %%
%%                                          %%
%% Please refer to the instructions for     %%
%% authors on:                              %%
%% http://www.biomedcentral.com/info/authors%%
%% and include the section headings         %%
%% accordingly for your article type.       %%
%%                                          %%
%% See the Results and Discussion section   %%
%% for details on how to create sub-sections%%
%%                                          %%
%% use \cite{...} to cite references        %%
%%  \cite{koon} and                         %%
%%  \cite{oreg,khar,zvai,xjon,schn,pond}    %%
%%  \nocite{smith,marg,hunn,advi,koha,mouse}%%
%%                                          %%
%%%%%%%%%%%%%%%%%%%%%%%%%%%%%%%%%%%%%%%%%%%%%%

%%%%%%%%%%%%%%%%%%%%%%%%% start of article main body
\section{Introduction}

In the recent years, several methods have been developed to gather quantitative data on human interactions using wearable sensors and complement more traditional methods based on surveys \cite{Mossong:PLOS2008,Danon:2013,Eames:2015}. Current data collection methods range from the use of Bluetooth or WiFi signals in mobile phones \cite{Eagle:PUC2006,Oneill:UbiComp2006,Scherrer:CN2008,Vu:MSWIM2010,Zhang:EPL2012,Stopczynski:PLOS2014} to the specific development of dedicated sociometric sensors \cite{Olguin:AMCIS2008,Hashemian:ISMOM2010,Salathe:PNAS2010,Cattuto:PLOS2010,Berke:AFM2011,Lucet:2012,Hornbeck:JID2012,Lowery-North:2013,Toth:2015,Guclu:PLOS2016} and enable researchers to record and measure physical proximity events between individuals in various social contexts. Depending on the specific technology considered however, spatial resolution varies and the resulting ``contacts'' detected can range from co-presence in a room or a part of a building to close face-to-face encounters. The resulting data is often temporally resolved and has been increasingly used in various contexts including the study of human behaviour, the validation of models of human interactions and data-driven models of epidemic spreading \cite{Blower:2011,Barrat:2014,Eames:2015}.

Despite the increasing availability of techniques to measure even high-resolution temporal contact networks however, a number of limitations remain. In particular, measures cannot be carried out for arbitrarily large population sizes. It is thus of crucial interest to infer contacts or build contact proxies from data with lower spatial resolution data or coming from other sources. In this spirit, several studies have considered the issue of inferring social ties from email exchanges \cite{DeChoudhury:ICWWW2010}, mobile phone data \cite{Eagle:PNAS2009}, or co-location at geographic scale \cite{Crandall:PNAS2010}. Other works try to infer close proximity in specific settings from individual attributes \cite{Scholz:ICWSM2013} or from a very precise localisation of individuals \cite{Lowery-North:2013}, or, at geographical scale,  from the similarity of the WiFi signals received from a large enough number of WiFi routers \cite{Sapiezynski:2017}.

Here instead, we do not try to infer specific contacts between pairs of individuals but rather investigate if a coarse co-location information on individuals allows us to reach an overall picture of the contact patterns in the population of interest. To this aim, we leverage several data sets collected by the SocioPatterns collaboration \cite{SocioPatterns,Cattuto:PLOS2010} in various contexts: these data include both detailed information about close, face-to-face encounters between individuals and a location tracking of individuals with low spatial resolution. It is thus possible to build two temporal networks where nodes represent individuals and links correspond respectively to a face-to-face contact or to a co-presence event, where co-presence is defined with respect to the localisation of two individuals within the same spatial area. We first compare the structural and statistical properties of these two temporal networks and show that they share some important properties, although the co-presence network is much denser, due to the lower spatial resolution involved in its definition. We thus investigate several methods of down-sampling the co-presence signal in order to create surrogate contact networks, in the spirit of \cite{Genois:NatComm2015,Sapienza:2017}, and compare these surrogate data to the actual networks of face-to-face contacts. We focus first on several statistical characteristics of temporal and aggregated networks, and quantify the ability to identify central nodes in the contact network from the surrogate data. We then consider the possibility to use the surrogate data in numerical simulations of data-driven models for epidemic spread. In particular, we compare the outcome of simulations of a standard model of epidemic propagation when surrogate or actual contact data are used, and we explore the possibility to identify the most efficient containment strategies from this limited information  \cite{Smieszek:BMC2013}. Our results turn out to depend strongly on the data collection context, highlighting the limitations of coarse co-presence networks with respect to detailed face-to-face data.

\section{The co-presence network}

\subsection{Data sets}

We use data collected by the SocioPatterns collaboration in various contexts. These data were gathered using wearable sensors able to detect face-to-face close range proximity (1.5\,m) of participants wearing the sensors on their chests. In addition, the sensors broadcast a signal that can be received by RFID readers located in the environment. In open space, each reader can receive signals from sensors situated within a range of $\sim 30\,$m, while the actual reception range in a building depends on its specific structure and on the nature of its walls, floors and ceilings. Each reader thus defines a coarse spatial area and the sensors' signals can be followed when the individuals carrying them change area. For each sensor, we define its ``spatial location'' at each time as the set of readers receiving its broadcasted signal at this time, and we define two individuals to be in co-presence if they share the same spatial location, i.e., the same exact set of readers have received signals from both individuals.

\begin{table}[h!]
  \caption{Characteristics of the data sets.}
  \label{tab:data}
  \begin{tabular}{|c|c|c|c|c|c|c|}
    \hline
    \bf Data set & \bf Location & \bf Year & $N_p$ & $N_a$ & $T$ & \bf Ref \\
    \hline
    InVS13      & Fr. Health Obs. & 2013 &  92 & 27 & 2 weeks & \cite{Genois:NWS2015}   \\
    InVS15      & Fr. Health Obs. & 2015 & 232 & 45 & 2 weeks & \\
    LH10        & Hospital        & 2010 &  81 &  8 & 3 days  & \cite{Vanhems:PLOS2013} \\
    LyonSchool  & Primary school  & 2009 & 242 & 15 & 2 days  & \cite{Stehle:PLOS2011}  \\
    SFHH        & Conference      & 2009 & 403 & 12 & 2 days  & \cite{Isella:JTB2011}   \\
    Thiers13    & High school     & 2013 & 326 & 18 & 1 week  & \cite{Fournet:PLOS2015} \\
    \hline
  \end{tabular}

  $N_p$ is the number of participants, $N_a$ the number of RFID readers, $T$ the total duration of the data collection.
\end{table}

We use data sets from various social contexts: a workplace, with data collected in two different years (InVS13, InVS15), a hospital (LH10), a primary school (LyonSchool), a scientific conference (SFHH) and a high school (Thiers13), see Table \ref{tab:data}.  In each case, we thus consider a temporal network of face-to-face contacts and a temporal network of co-presence between individuals, both at the temporal resolution of $20 s$. A contact (resp. co-presence) event between two individuals is then defined as a set of successive time-windows of $20 s$ during which the individuals are detected in contact (resp. co-presence), while they are not in the preceding nor in the next $20 s$ time window. While the conference data does not include any other information on the participants and does not exhibit any particular group structure \cite{Stehle:BMC2011}, the other populations under study can each be divided into groups: departments for the workplace, classes for the school and the high school, and roles (patients, doctors, nurses) in the hospital. In these cases, the overall structure of networks aggregated over a certain time window can be summarised, in addition to usual quantities such as the density, the clustering coefficient or the degree distribution, by contact (resp. co-presence) matrices that give the fraction of pairs of individuals of different groups who have been in contact (resp. in co-presence). Moreover, temporal features of interest include the distributions of durations of contact or co-presence events, of the time elapsed between successive events, of the numbers and aggregated durations of such events between pairs of individuals (the latter quantity yields a natural definition of the weight of a link between individuals in the aggregated network).

We will show in the main text the results corresponding to the InVS15 data set, and we refer to the Supplementary Information for the results obtained with the other data sets. We make also available as Supplementary Files the temporally resolved contact and co-presence networks.

\subsection{Co-presence and contact networks}

\begin{figure}[h!]
  \includegraphics[width=.7\columnwidth]{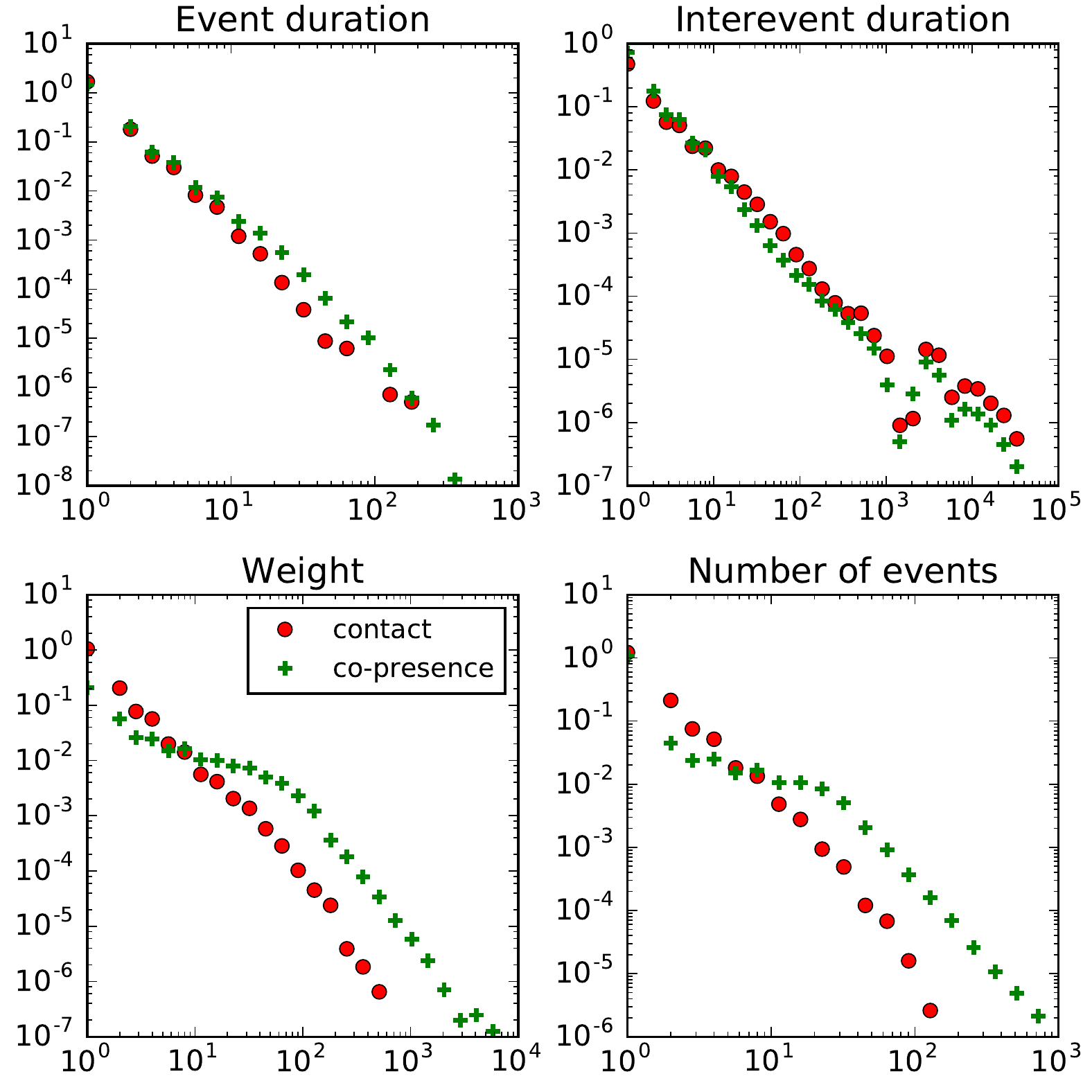}
  \caption{\csentence{Temporal distributions for co-presence and contact events --- InVS15.} We show for both the contact and co-presence of the same data set the distributions of event and inter-event duration, link weights (as total contact duration) and number of contacts per link.}
  \label{fig:distrib_InVS15}
\end{figure}

We first compare some features of the co-presence and contact networks, both temporal and for networks aggregated either on the whole data gathering period or over daily temporal windows. We show in Fig.~\ref{fig:distrib_InVS15} the distributions of event and inter-event duration, as well as the distributions of number and cumulative duration of events for individual pairs. The co-presence events show broad distributions of these quantities, similarly to the contact events and with similar slopes: using only co-presence data yields approximate information on the functional shape of the contact duration distributions. The distributions of durations and numbers of events are however typically broader for co-presence, with heavier tails, and the distribution of inter-event durations tend to be less broad (see also SI). This is not surprising as the criterion for being in co-presence is less strict than for being in contact. We observe the strongest differences between co-presence and contact distribution functional shapes for the primary school data. This could be related by the fact that the spatial resolution is in that case quite low, with all the schoolyard being covered by one single reader, and some readers covering more than one classroom. Overall, using only co-presence data would lead to over-estimations of the contact durations and aggregate durations.

\begin{table}[h!]
  \caption{Similarity between contact matrices.}
  \label{tab:sim_cmco}
  \begin{tabular}{|c|c|c|c|c|c|}
    \hline
    & InVS13 & InVS15 & LH10 & LyonSchool & Thiers13 \\
    \hline
    Co-presence & 0.790 & 0.710 & 0.968 & 0.706 & 0.681 \\
        \hline
    Sampling 1 & 0.946 & 0.829 & 0.960 & 0.845 & 0.857 \\
    Sampling 2 & 0.958 & 0.901 & 0.894 & 0.945 & 0.937 \\
    Sampling 3 & 0.888 & 0.816 & 0.958 & 0.738 & 0.691 \\
    \hline
  \end{tabular}

  For each data set we compute the cosine similarity between the average daily contact matrix and the co-presence matrix, as well as for the contact matrices obtained for each sampling method of the co-presence data, averaged over $100$ realisations for each sampling method.
\end{table}

\begin{table}[h!]
  \caption{Characteristics of the contact, co-presence, and sampled co-presence networks.}
  \label{tab:k_rho}
  \begin{tabular}{|c|c|c|c|c|c|c|}
    \hline
                  & InVS13 & InVS15 & LH10 & LyonSchool & SFHH & Thiers13 \\
    \hline
    $\bar{k}_c$   &  2.9 &  6.4 & 14.0 &  47.3 &  28.8 &  13.5 \\
    $\bar{k}_\ell$ & 20.9 & 35.0 & 18.2 & 194.5 & 234.3 & 126.8 \\
        \hline
    $\bar{k}_1$   &  5.8 & 14.2 & 14.4 & 101.3 & 116.7 &  52.2 \\
    $\bar{k}_2$   &  0.9 &  3.6 &  7.7 &  36.9 &  45.2 &   5.2 \\
    $\bar{k}_3$   &  5.3 &  5.0 & 14.0 &  21.2 &  40.4 &   4.1 \\
    \hline    \hline
    $\rho_c$   & 0.030 & 0.028 & 0.175 & 0.196 & 0.072 & 0.041 \\
    $\rho_\ell$ & 0.211 & 0.152 & 0.227 & 0.807 & 0.807 & 0.383 \\
        \hline
    $\rho_1$   & 0.058 & 0.061 & 0.179 & 0.420 & 0.290 & 0.158 \\
    $\rho_2$   & 0.009 & 0.016 & 0.097 & 0.153 & 0.112 & 0.016 \\
    $\rho_3$   & 0.054 & 0.022 & 0.175 & 0.088 & 0.101 & 0.013 \\
    \hline    \hline
    $\omega_c$   &  4.4 &  7.6 & 14.3 &  22.5 & 11.0 &  9.4 \\
    $\omega_\ell$ & 18.8 & 38.7 & 22.7 & 141.0 & *    & 74.6 \\
        \hline
    $\omega_1$   &  6.6 & 10.3 & 17.2 &  41.5 & 34.7 & 33.8 \\
    $\omega_2$   &  3.0 &  5.3 &  8.6 &  12.8 & 12.2 &  3.9 \\
    $\omega_3$   &  5.5 &  4.8 & 17.1 &   6.3 &  9.0 &  3.8 \\
    \hline    \hline
    $\bar{c}_c$   & 0.178 & 0.239 & 0.428 & 0.520 & 0.260 & 0.379 \\
    $\bar{c}_\ell$ & 0.417 & 0.409 & 0.491 & 0.868 & 0.880 & 0.581 \\
        \hline
    $\bar{c}_1$   & 0.255 & 0.266 & 0.432 & 0.596 & 0.442 & 0.586 \\
    $\bar{c}_2$   & 0.045 & 0.139 & 0.309 & 0.370 & 0.212 & 0.092 \\
    $\bar{c}_3$   & 0.205 & 0.101 & 0.426 & 0.193 & 0.161 & 0.047 \\
    \hline
  \end{tabular}

  We compare the average degree ($\bar{k}$) network density ($\rho$), clique number ($\omega$) and average clustering ($\bar{c}$) of daily aggregated networks, for the contact network ($c$ subscript), the co-presence network ($\ell$ subscript), and the sampled co-presence networks (subscripts $1$ to $3$ according to the sampling method). Values are averaged over all the days of the study. In the case of SFHH, since on the second day there was activity only during the morning, only the values of the first day are reported.\\ *The network is too large and too dense for the clique number to be determined in reasonable time via the usual algorithm.
\end{table}

\begin{figure}[h!]
  \includegraphics[width=\columnwidth]{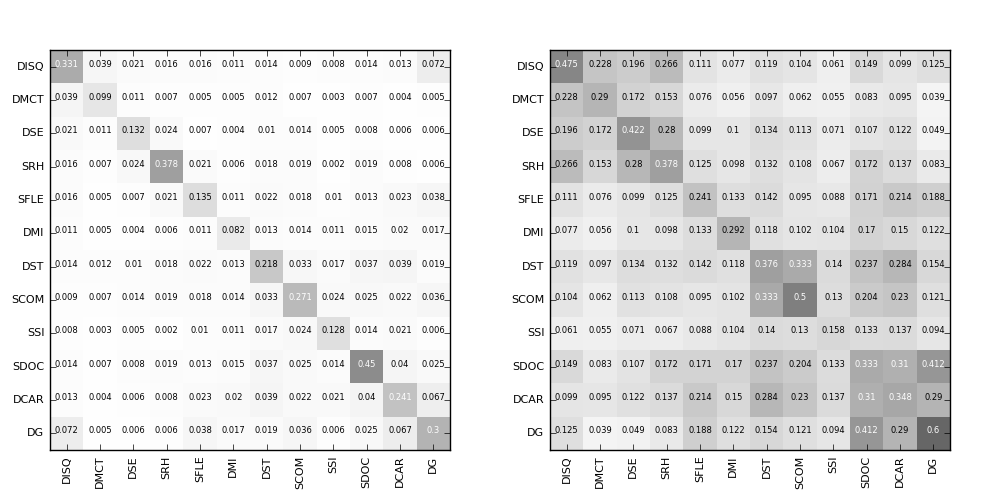}
  \caption{\csentence{Contact and co-presence matrices --- InVS15.} Comparison between the average matrices of link density for the contacts and the co-presence daily aggregated networks. Values are averaged over all days of the data collection. Both plots have the same colour scale.}
  \label{fig:CM_InVS15}
\end{figure}

\begin{figure}[h!]
  \includegraphics[width=.8\columnwidth]{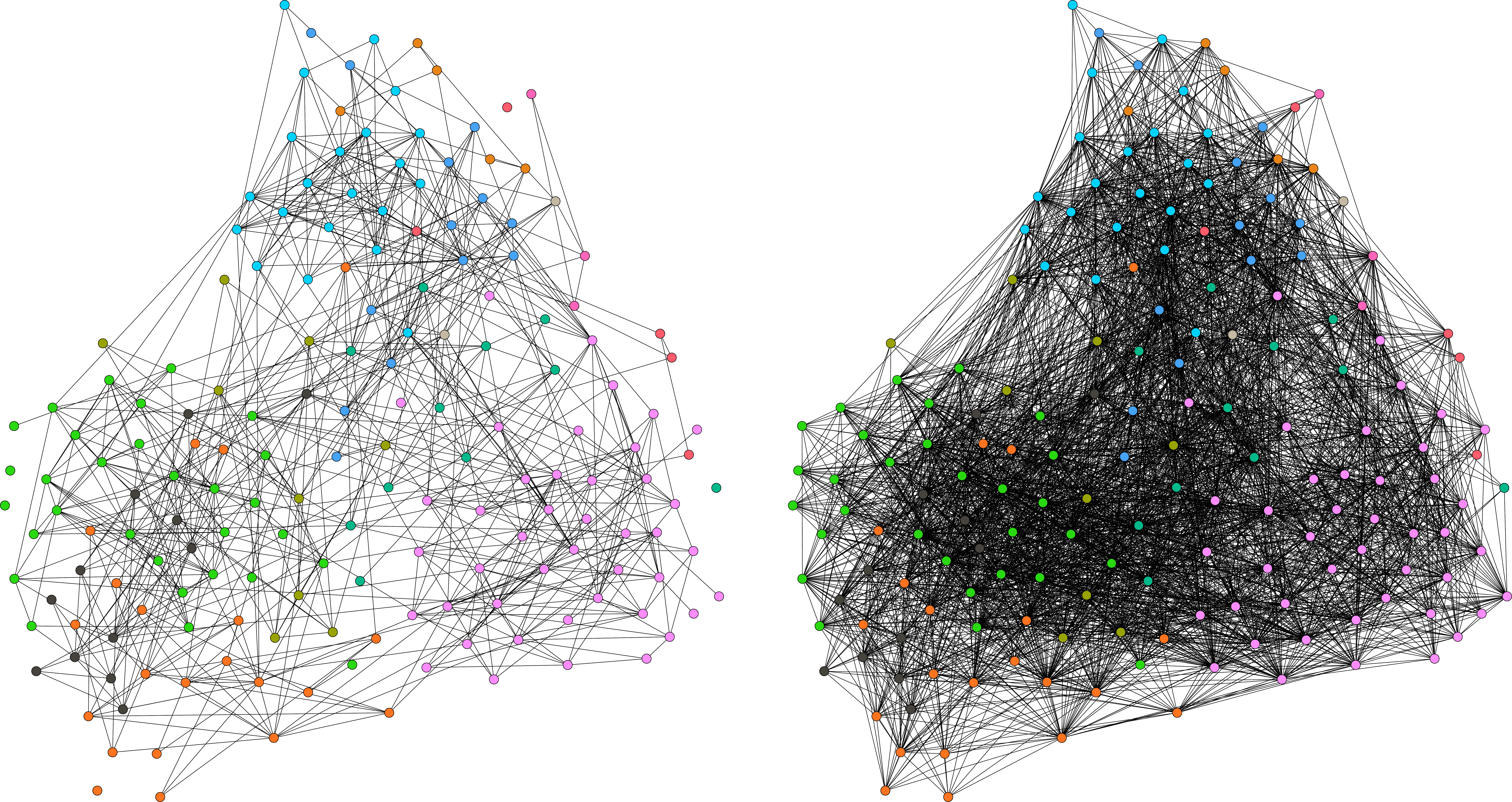}
  \caption{\csentence{Day 2 contact and co-presence networks --- InVS15.} This figure highlights the difference in terms of link density when comparing contact and co-presence daily networks. Different node colours correspond to the different administrative departments.}
  \label{fig:graph_InVS15}
\end{figure}

We compare moreover in Figs.~\ref{fig:CM_InVS15} - \ref{fig:graph_InVS15} and Tables \ref{tab:sim_cmco} - \ref{tab:k_rho} the overall structures of the contact and co-presence networks, aggregated over daily time windows. The co-presence aggregated networks are much denser than the contact network, with a larger average degree, a larger average clustering coefficient and larger cliques, as expected once again given the lower spatial resolution required for co-presence events. In some cases (school, conference), the aggregated networks are even close to being fully connected (see for illustration Fig.~\ref{fig:graph_InVS15}). Despite this strong difference in the overall density of links, the contact and co-presence matrices giving the density of links between and within each group, averaged across days, are very similar (Table \ref{tab:sim_cmco}). The similarity is particularly high for the hospital data and, even for the lower value obtained for the high school data, the matrices displayed in the SI show that the overall structure in classes and groups of classes can be inferred from the co-presence data alone.

\begin{figure}[h!]
  \includegraphics[width=.5\columnwidth]{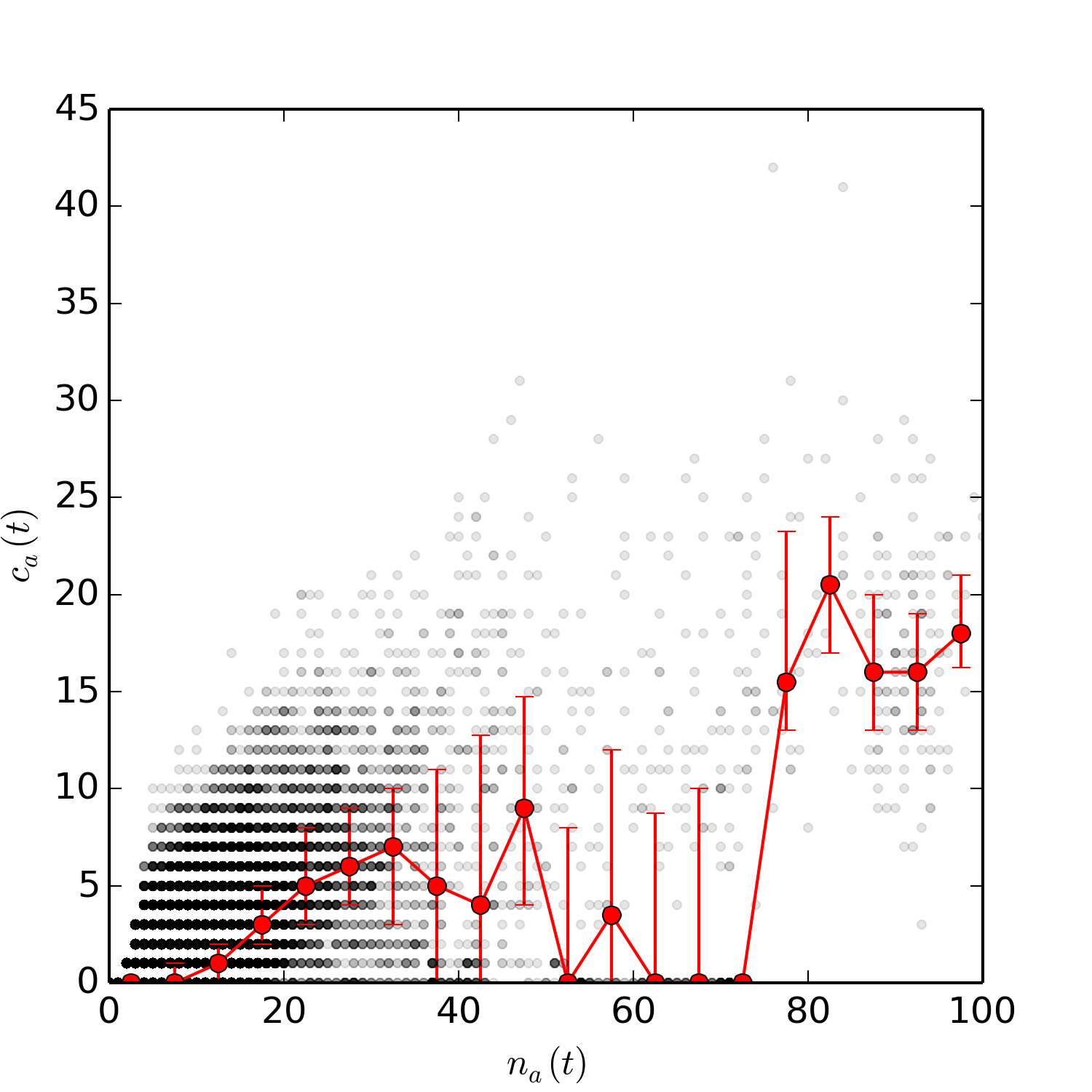}
  \caption{\csentence{Number of contacts as a function of the number of individuals present --- InVS15.} We plot the number of contacts $c_a(t)$ occurring at time $t$ in a certain area $a$ as a function of the number of individuals $n_a(t)$ present at the same time in $a$. The red line shows the median of the scatter plot, with error bars defined by the $25\,\%$ and $75\,\%$ percentiles.}
  \label{fig:TL_InVS15}
\end{figure}

Given the simultaneous discrepancies in density values and similarities in the networks group structures, we investigate if the data exhibits a scaling law between the number of individuals present in an area and their contact activity, as found at geographical scale in phone communication \cite{Schlapfer:2014} and Twitter data \cite{Tizzoni:2015}. Figure~\ref{fig:TL_InVS15} and the similar figures shown in SI show the results obtained in the various contexts. Apart from the office cases (InVS13 and InVS15), we observe indeed a correlation between the median of the number of contacts and the number of individuals present. This correlation exhibits a power law shape, with an exponent around 1.5 (see figures in SI). However, huge, context-dependent fluctuations are observed. For instance, in the InVS15 case, the trend is strongly influenced by the numerous instances of an absence of contacts despite potentially large values of the number of individuals present in the area. This is a consequence of the fact that a given reader can receive signals from the sensors of individuals located in different offices. In other areas such as a cafeteria, many more contacts occur with potentially a similar or even smaller number of individuals. Overall, very large fluctuations of the number of contacts, at given number of individuals present, are thus observed, because on the one hand of the low spatial resolution of the co-presence data, and on the other hand of the variety of contexts corresponding to the areas covered by different RFID readers. The stronger correlation is observed for the SFHH conference data, probably because the various areas covered by the readers corresponded to similar contexts, namely different areas of the exhibition and poster rooms.

\section{Sampling co-presence data}

\subsection{Sampling methods}

As the temporal network of co-presence bears some similarities with the actual contact data, but contains much more events and leads to much denser aggregated networks, we consider the possibility to down-sample the co-presence data: for each pair of individuals, each contact event is indeed included in a co-presence event of the same individuals. Each co-presence event might thus correspond to one or more contact events. As we cannot determine exactly the correct down-sampling to be performed if we have access only to co-presence data, we study here three simple sampling methods. We remind here that we do not try to infer the real contacts but rather to obtain a down-sampled version of the co-presence network that is statistically similar to the real contact data. Moreover, as the total number and duration of actual contacts cannot either be easily guessed from the co-presence data alone, we consider the actual total contact time $T_c$ as the (only) parameter of the sampling, and we fix it to its empirical value. The sampling methods we consider are the following:

\begin{itemize}
\item {\bf Sampling 1: Sampling of co-presence times.} We define a co-presence list as a list of individuals present at the same time $t$ in the same area. Each co-presence list is thus stamped with its time of occurrence $t$. We create $n_\ell$ copies of each co-presence list $\ell$, where $n_\ell$ is the number of distinct individuals in $\ell$, and create in this way of a global pool of co-presence lists. We then sample $T_c$ lists uniformly at random from the pool without replacement. Each list has thus a probability proportional to the number of individuals it contains to be chosen. From each chosen list, we choose at random a pair $i,j$ of individuals, obtaining  a triplet $(t,i,j)$ where $t$ is the time-stamp of the list (we take care to avoid repetitions: if $(t,i,j)$ has already be obtained in a previous random draw, we repeat the random selection). The sampled temporal co-presence network (i.e., the surrogate contact network) is formed by the union of these triplets.

\item {\bf Sampling 2: Sampling of co-presence times with completion.} 
We constitute a pool of lists exactly like in the previous method. We then sample a triplet $(t,i,j)$ as in the previous method, and add all the other triplets $(t',i,j)$ that belong to the same co-presence event to create the surrogate contact event. We iterate this until we reach a cumulative contact time $T_c$, while discarding repetitions.

\item {\bf Sampling 3: Sampling of co-presence events.} 
We consider directly the list of co-presence events between individuals, $(t,i,j,\tau)$ (co-presence event between individuals $i$ and $j$,  starting at time $t$ and with duration $\tau$), and sample events from this list, without replacement, adding them to the list of surrogate contact events until we reach a cumulative contact time $T_c$.
\end{itemize}

For each data set, we create $100$ instances of surrogate contact networks for each sampling method. We compare in the following the properties of these surrogate contact networks with the real face-to-face contact data.

\subsection{Network comparison}

\begin{figure}[h!]
  \includegraphics[width=.9\columnwidth]{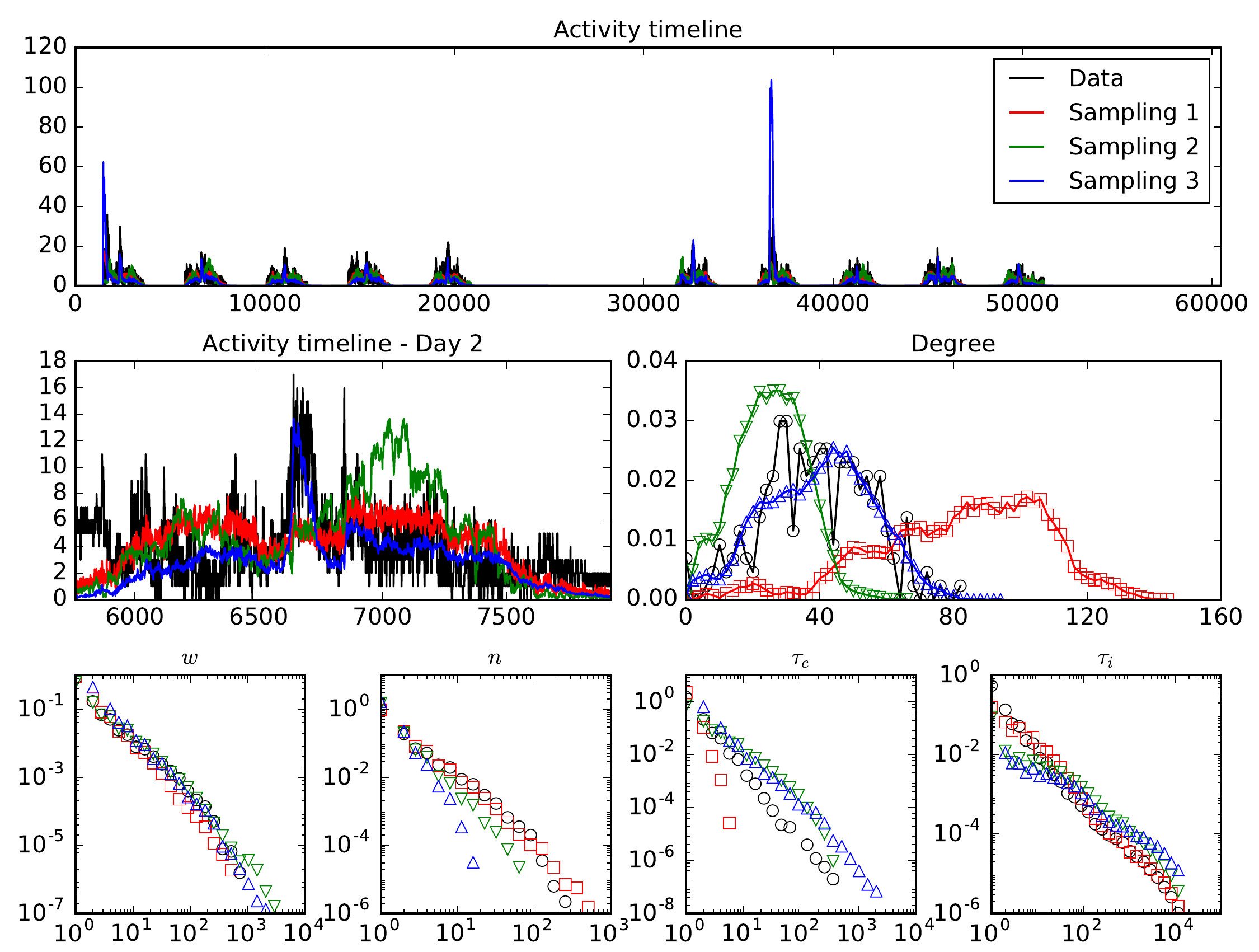}
  \caption{\csentence{Properties of the sampled co-presence networks --- InVS15.} We compare several properties of the contact network from the original data set with the surrogate contacts obtained by sampling of the co-presence data: overall timeline of contact activity, distributions of degree, weight $w$ and number of contacts per link $n$ in the network aggregated over the whole data collection period, and distributions of the contact duration $\tau_c$ and inter-contact duration $\tau_i$.}
  \label{fig:analyse_InVS15}
\end{figure}

\begin{figure}[h!]
  \includegraphics[width=.5\columnwidth]{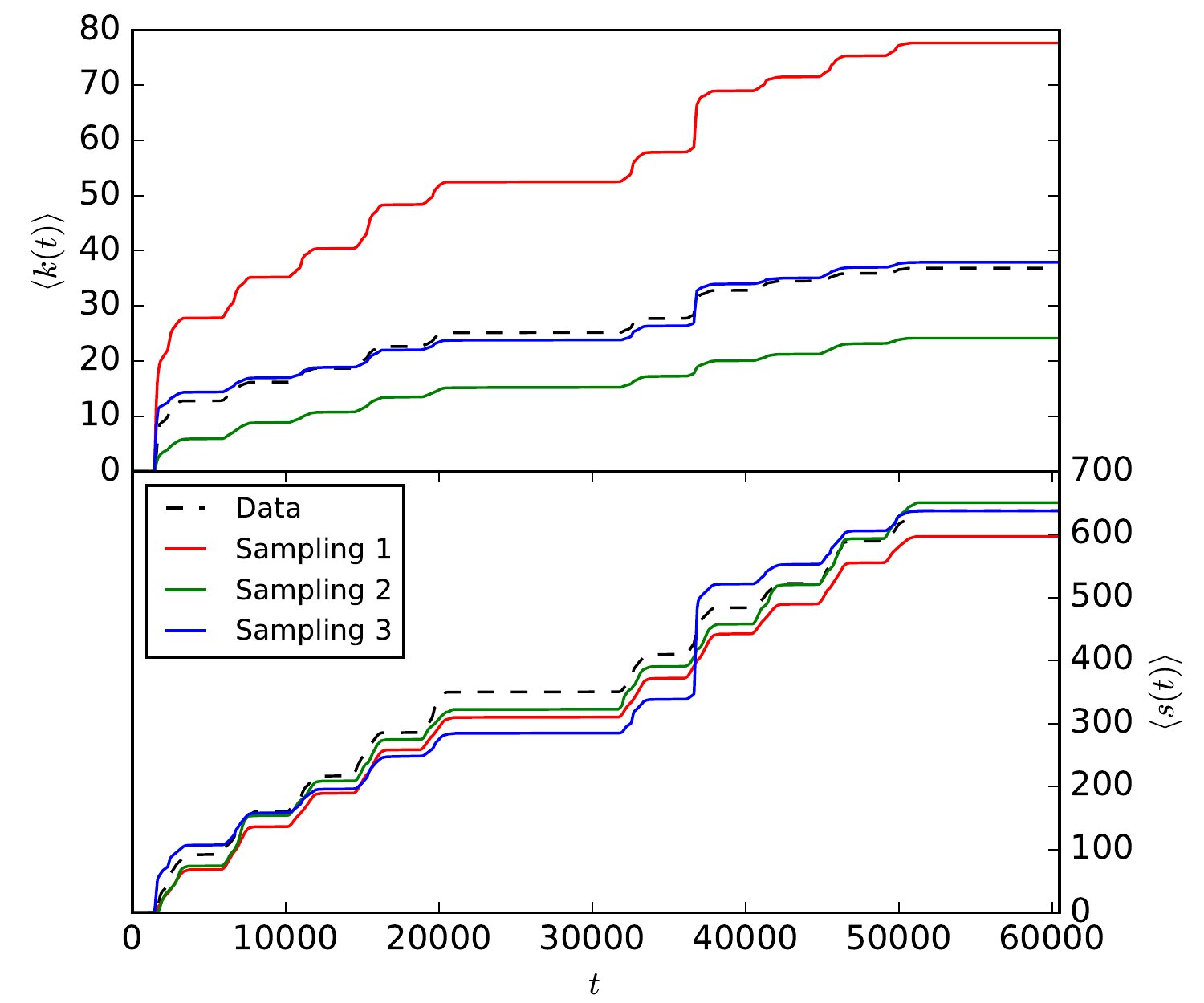}
  \caption{\csentence{Evolution of the mean aggregated degree and strength. InVS15 data set} We compare how the average degree $\left<k(t)\right>$ and the average strength $\left<s(t)\right>$ grow as we aggregate the network on increasing time-windows, for the real contact data and each sampling method.}
  \label{fig:tl_ks_InVS15}
\end{figure}

Figures \ref{fig:analyse_InVS15} - \ref{fig:tl_ks_InVS15} and Tables \ref{tab:sim_cmco} - \ref{tab:k_rho} provide elements of comparison between the surrogate contact networks and the empirical data (see also SI). The first observation is that the contact activity timelines are in general broadly recovered, except for the primary school (see SI), while the detailed intra-day activity variations are not always properly reconstructed in the surrogate data (except for the hospital data, see SI). The strongest deviations are observed for the second sampling method for the conference and high school data.

The first sampling method, given it samples separately times of co-presence, yields an exponential distribution of surrogate contact duration, in contrast with actual data and other sampling methods in which broad distributions are observed. Broad distributions of inter-contact durations and of the numbers of contacts between individuals are also obtained, with however slopes that depend on the context. For instance, the second sampling method systematically leads to a distribution of contact durations that is broader than for the real contacts. The third method yields a distribution of contact durations similar to the real one for the InVS13, LH10, and SFHH cases, but gives results similar to the second method in the other cases.

We now turn to the properties of networks aggregated over daily periods or over the whole data collection. At the daily level, we show in Table \ref{tab:sim_cmco} that the similarity of the contact matrices obtained from the surrogate data with the empirical one is very high, and most often larger than the similarity of the original co-presence matrix. For networks aggregated over the whole data collection, Fig.~\ref{fig:analyse_InVS15} shows  the distributions of degrees and of weights (see also SI). The first sampling method leads to an overestimation of degree values (resulting in a shift of the distribution), the second method tends to shift the distribution to lower degree values (except for the conference case), and the third method yields context-dependent over- or under-estimations of degree values. Note that the distributions of degrees of the co-presence networks are not shown in the figure as the degree values are very strongly overestimated. Distributions of weights (aggregated contact durations) recover well the ones of the data for all sampling methods, and are closer than the ones of the co-presence networks.

\begin{table}[h!]
  \caption{Average similarity between daily networks.}
  \label{tab:sim_net}
  \begin{tabular}{|c|c|c|c|c|c|c|}
    \hline
                & InVS13 & InVS15 & LH10  & LyonSchool & Thiers13 \\
    \hline
    Contact     & 0.333  & 0.305  & 0.351 & 0.643      & 0.431    \\
    Co-presence & 0.415  & 0.348  & 0.449 & 0.806      & 0.683    \\
    Sampling 1  & 0.361  & 0.344  & 0.436 & 0.749      & 0.515    \\
    Sampling 2  & 0.388  & 0.271  & 0.403 & 0.175      & 0.084    \\
    Sampling 3  & 0.286  & 0.205  & 0.437 & 0.042      & 0.071    \\
    Null model  & 0.022  & 0.010  & 0.061 & 0.046      & 0.010    \\
    \hline
  \end{tabular}

  For each data set we compute the cosine similarity between the neighbourhoods of each nodes from each daily network, averaged for all nodes and all pairs of daily networks. The neighbourhood of a node $n$ is defined as the vector of the link weights between $n$ and every other nodes (if the link does not exist the weight is set to zero). We compare the values obtained for the contact data, the co-presence data, and for the networks generated by each sampling method of the co-presence data, averaged over $100$ realisations for each sampling method. For reference, we also compute as null model the average similarity when links in the contact data are shuffled randomly within each daily network.
\end{table}

To investigate intermediate timescales of aggregation, Table \ref{tab:sim_net} quantifies the similarity between networks aggregated in different days. The measure is defined as the average cosine similarity between all pairs of instances of a node's neighbourhood, averaged over all nodes. We see that the similarity is higher for the co-presence networks, as expected since the networks are denser. The sampling method 1 generates networks that are more similar than the data, and the other two methods generate networks that are less similar (with the exception of the LH10 case, and the method 2 in the InVS13 case). In the cases of the method 2 for the LyonSchool data, and the methods 2 and 3 for the Thiers13 data, the sampled networks are even almost as different as they would be after a random shuffling of the links.

In addition, Fig.~\ref{fig:tl_ks_InVS15} gives the evolution of the average degree and strength for networks aggregated in increasingly long time windows. First, the evolution of the real average aggregated strength is usually better recovered than for the degree by the various sampling better. Second, which sampling method recovers better the evolution of the degree is again context dependent. However, in all cases the sampled data are much closer to the contact data than the co-presence network, which overestimates very strongly these quantities.

\subsection{Node centralities}

\begin{figure}[h!]
  \includegraphics[width=.7\columnwidth]{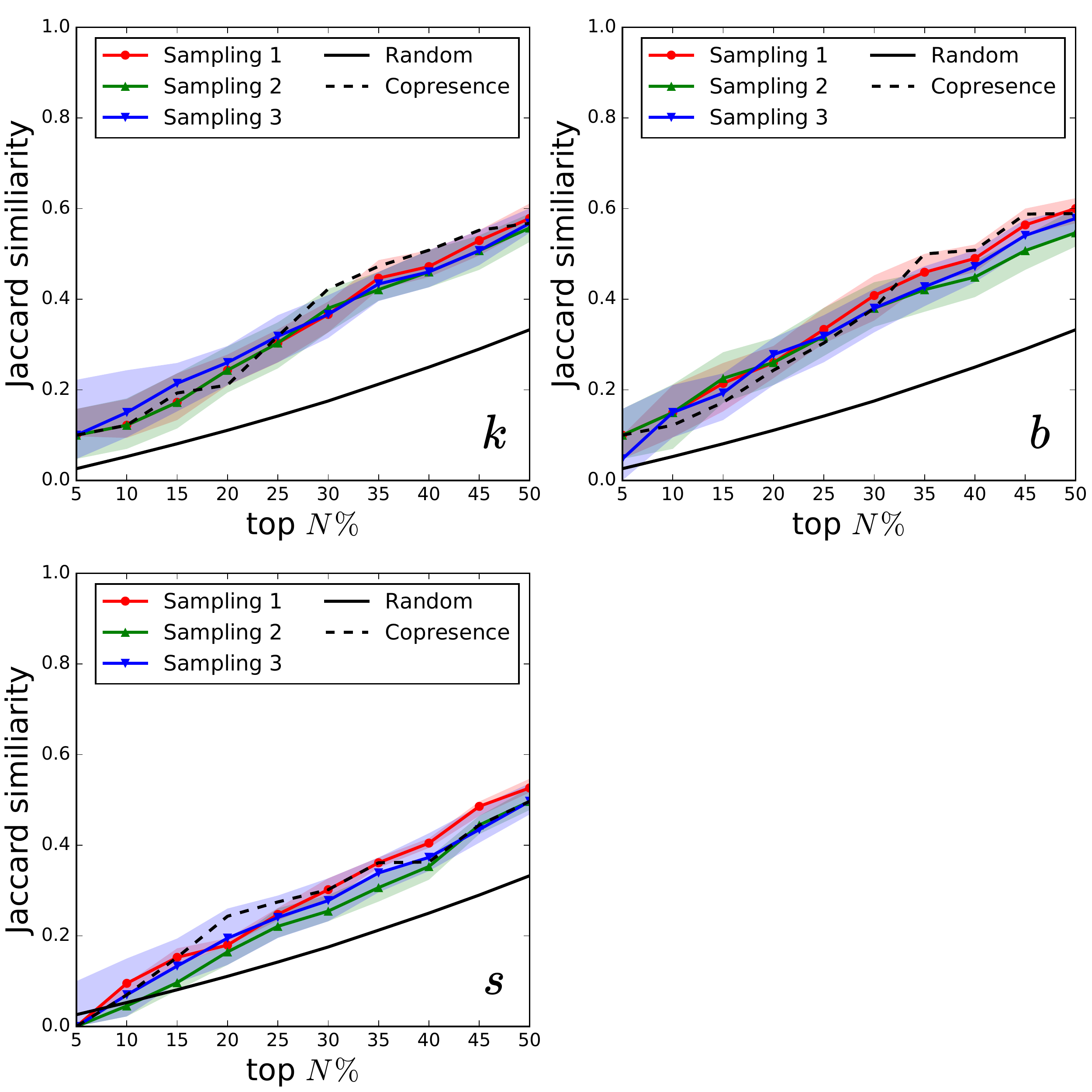}
  \caption{\csentence{Node ranking similarity. InVS15 data set} We plot for each co-presence sampling method the Jaccard similarity between the top $N$\,\% nodes in the real and surrogate contact data, when ranked according to their degree $k$, their strength $s$ or their betweenness centrality $b$ vs. $N$. The plot shows the median similarity and the shaded areas give the $90\,\%$ confidence interval.}
  \label{fig:rank_InVS15}
\end{figure}

In a network, more ``central'' nodes are usually considered as important, as they might play an important role for instance in spreading processes (or other dynamical phenomena) occurring in the network. It is thus of interest to understand whether the most central nodes in the contact network can be identified either in the raw co-presence data or in the surrogate contact data built from the co-presence information. As there are several ways of determining central nodes in a network, we consider here three of the most well-known centrality measures and apply them to the networks aggregated over the whole data collection: degree $k$, strength $s$ and betweenness $b$ of nodes in the aggregated networks. For each instance of each sampling method, we thus build the resulting surrogate aggregated contact network and rank nodes according to each centrality measure. We then compute the Jaccard similarity index between the top $N\,\%$ nodes in the real contact network and in the surrogate one. We plot in Fig.~\ref{fig:rank_InVS15} the median similiarity with the $90\,\%$ confidence interval, as a function of $N$, for the InVS15 case (see SI for the other cases).

\begin{table}[h!]
  \caption{Comparison of the maximum k-core properties.}
  \label{tab:kcore}
  \begin{tabular}{|c|c|c|c|}
    \hline
                & InVS13       & InVS15       & LH10         \\
    \hline
    Contact     & 11           & 25           & 23           \\
    Co-presence & 78   (0.607) & 112  (0.681) & 32   (0.682) \\
    Sampling 1  & 22.5 (0.660) & 57.8 (0.719) & 26.0 (0.683) \\
    Sampling 2  & 5.23 (0.479) & 17.4 (0.692) & 16.4 (0.655) \\
    Sampling 3  & 28.1 (0.591) & 27.4 (0.639) & 25.5 (0.693) \\
    \hline
    \hline
                & LyonSchool   & SFHH         & Thiers13     \\
    \hline
    Contact     & 47           & 33           & 24           \\
    Co-presence & 181  (0.615) & 320  (0.522) & 210  (0.684) \\
    Sampling 1  & 99.8 (0.638) & 111  (0.575) & 76.7 (0.640) \\
    Sampling 2  & 39.7 (0.501) & 41.4 (0.555) & 17.7 (0.375) \\
    Sampling 3  & 28.4 (0.360) & 42.6 (0.559) & 15.8 (0.117) \\
    \hline
  \end{tabular}

For each dataset we compute the maximum coreness, and report between parenthesis the Jaccard index between the k-core of the contact network and the k-core in the original and sampled  co-presence data (results are averaged over $100$ realisations for each sampling method).
\end{table}

In general, no sampling method recovers correctly the most central nodes for low values of $N$. The best results are obtained for the conference data with similarities around $0.2 - 0.4$. The similarity values increase as $N$ increases but reach most often only values of $\sim 0.5$ when considering the top 50\,\% nodes, meaning that only 25\,\% of the most central nodes are identified when using the surrogate data. The best results are obtained for the first sampling method for the LyonSchool case and for the LH10 case, with similarities reaching $0.6  - 0.7$. Results are typically better than the random baseline but do not outperform the detection of most central nodes based on the whole co-presence network. In terms of the most central nodes as defined by the $k$-core decomposition (we recall that the $k$-core of a network is the maximal subgraph such that all nodes in the subgraph have at least degree $k$, and $k$ is called the coreness), the overestimation of degrees in the co-presence network leads to an overestimation of the maximum coreness, while sampling leads to values closer to the ones of the contact data, but once again in a context-dependent way. The maximum core itself is 
only partially recovered in the whole and in the sampled co-presence networks (see Table \ref{tab:kcore}).

\section{Using surrogate contact data in epidemic simulations}

We have seen in the previous section that none of the three sampling methods yields a perfectly accurate description of all the relevant features of the true contact network: each sampling method yields surrogate data with both interesting similarities and potentially important discrepancies with respect to the original contact data. We now consider the issue of using such surrogate data in simulations of spreading processes: as precise data on face-to-face contacts is not always available, it is important to understand if co-presence information can allow us to obtain on the one hand an accurate prediction of the outcome of an epidemic process, and on the other hand a reliable estimation of the impact of containment measures. In particular, it is important to be able to classify potential containment strategies to determine which one(s) are most adequate.

To this aim, we consider the paradigmatic Susceptible-Infectious-Recovered (SIR) model for epidemic spreading. In this model, susceptible (S) individuals can become infectious (I) at rate $\beta$ when in contact with an infectious node. Infectious nodes recover spontaneously at rate $\mu$ and enter an immune {\sl recovered} (R) state. Simulations start with a single infectious individual chosen at random and carried out until there are no infectious individuals left in the population, i.e., individuals are either still susceptible or have been infectious and have then recovered. The impact of the epidemics is then quantified by the final fraction $n_i$ of individuals in the R state. 

We set $\beta = 0.0004$ and vary $\mu$ by tuning the reproductive number $R = \beta/\mu$. For each value of $R$, we measure the fraction $P(n_i > 20\,\%)$ of ``large'' outbreaks in which the fraction $n_i$ of the population that was reached by the outbreak is at least $20\,\%$ and the distribution of the sizes $n_i$ of these large outbreaks. We average the results over $10\,000$ simulations performed on the empirical contact network. For each sampling method, we build $100$ different instances of the surrogate contact network, and perform $100$ simulations on each surrogate network. 
 
We also consider several simple methods to mitigate the spread, namely the vaccination of a number of individuals in the population, under the assumption of a perfect vaccine efficiency: vaccinated individuals cannot become infectious nor transmit the disease and thus slow down and hinder the propagation. We consider the vaccination of (i) $5$, $10$ or $20$ individuals chosen at random (ii) the most central $5$, $10$ or $20$ individuals, where centrality is measured according to either degree, strength or betweenness in either the real or surrogate contact networks (iii) when the population is structured in groups, the vaccination of all individuals in one group.

\begin{figure}[h!]
  \includegraphics[width=.5\columnwidth]{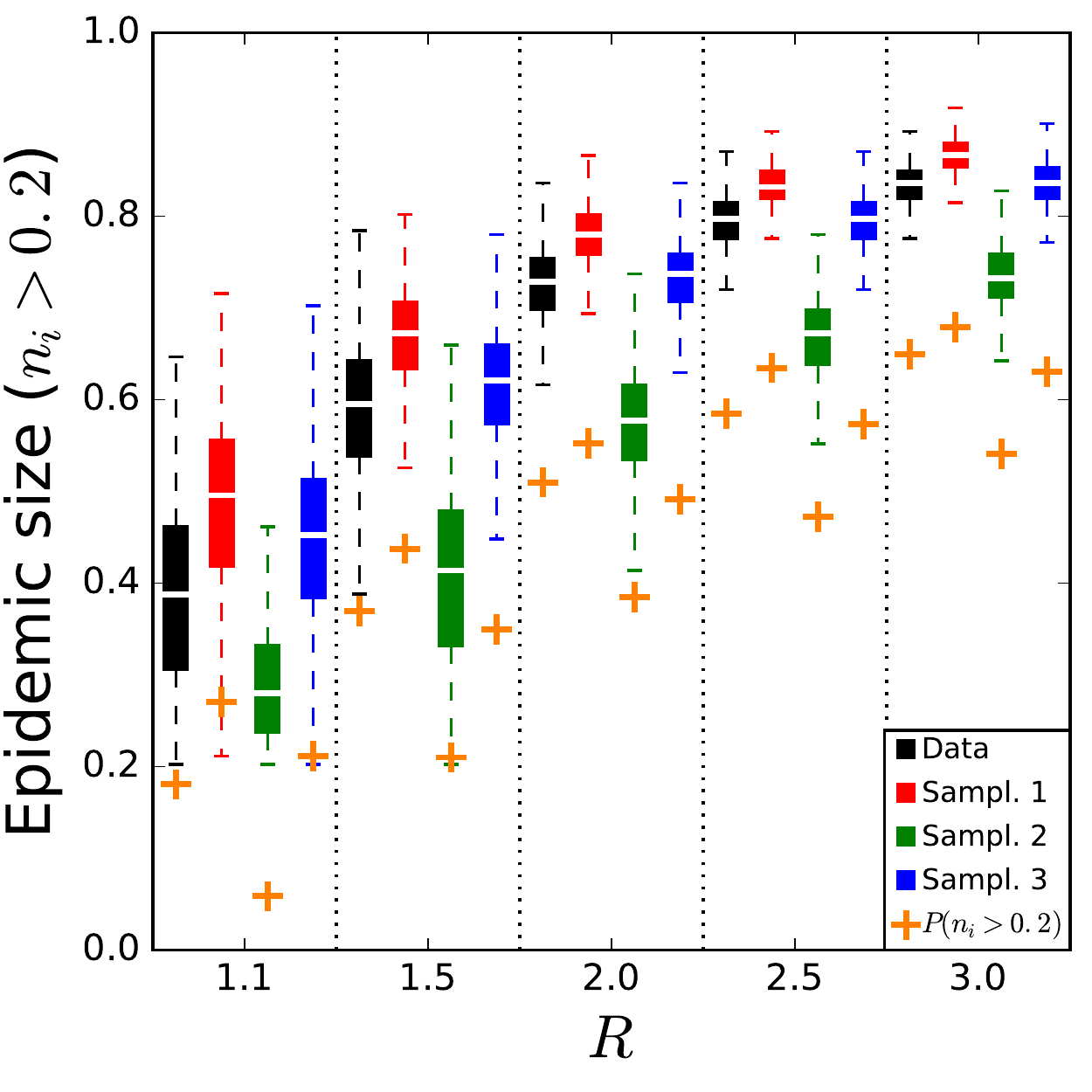}
  \caption{\csentence{Epidemic prevalence. InVS15 data set.} We plot the fraction of the total number of outbreaks that reach at least 20\,\% of the population (crosses) and the distribution of the sizes of these outbreaks (boxplots) for several values of the reproductive number $R$, for the original  and the surrogate contact networks.}
  \label{fig:epidemic_InVS15}
\end{figure}

\begin{figure}[h!]
  \includegraphics[width=.5\columnwidth]{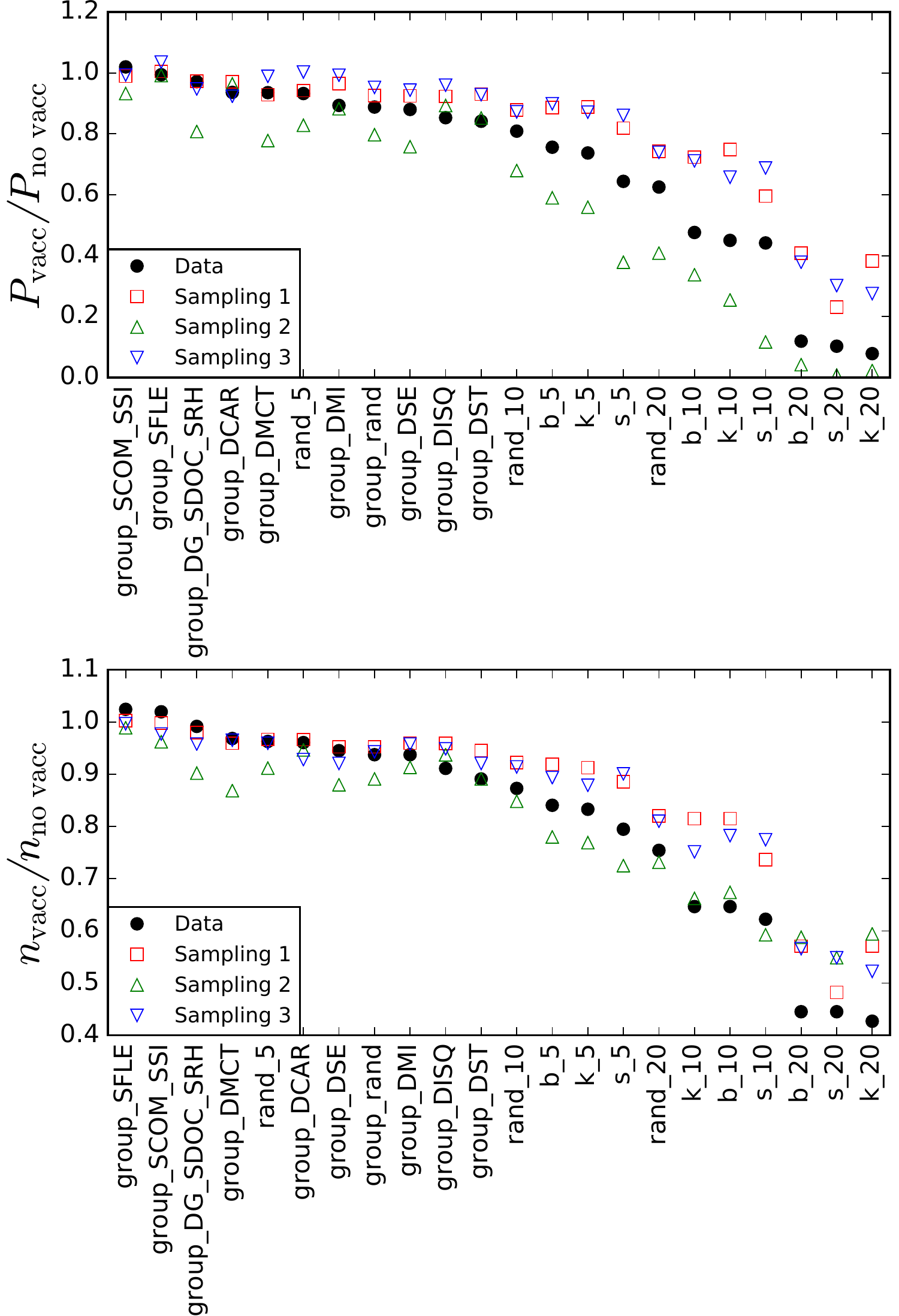}
  \caption{\csentence{Vaccination strategies. InVS15 data set.} We plot the ratio between the vaccination and no vaccination cases of the fraction of the total number of outbreaks that reach at least 20\,\% of the population (top), and of the median size of these outbreaks (bottom) for different vaccination strategies, for the original data and the reconstructed networks. The vaccination strategies are ordered by decreasing efficiency, based on the effect on the real contact data. The \emph{group\_*} strategies consist in vaccinating one or several groups entirely; the \emph{group\_rand} strategy vaccinates $n_g$ random nodes, where $n_g$ is the average group size; the \emph{rand\_n} strategies randomly vaccinates a specified fraction $n$ of nodes; the \emph{b\_n}, \emph{k\_n}, \emph{s\_n} strategies vaccinate the top $n$\,\% nodes according to, respectively, betweenness centrality, degree and stregnth ranking.}
  \label{fig:vacc_InVS15}
\end{figure}

Figures \ref{fig:epidemic_InVS15} and \ref{fig:vacc_InVS15} summarize our results for the InVS15 dataset (see SI for the figures obtained with the other datasets). In terms of the evaluation of the impact of a spreading process, results are context dependent. The simulations performed on the surrogate data obtained with the first method generally lead to an overestimation of the epidemic risk, except for the hospital data. When using the second sampling method, we obtain a good estimation of the risk for the conference, school and highschool data but an underestimation for offices and hospital data. The third method on the other hand leads to a correct estimation for the offices and hospital data but an underestimation for the school and highschool and an overestimation for the conference.
 
We show in Fig.~\ref{fig:vacc_InVS15} the impact of the various vaccination strategies, quantified through the ratio of the probabilities of large outbreaks with and without vaccination, as well as the ratio between the median sizes  of these large outbreaks. We rank the strategies according to their efficiency in the real contact network, in order to visualize easily whether the surrogate networks lead to the same classification of the strategies: indeed, even when the impact of each specific strategy is not accurately quantified, it would be interesting at least to understand which methods are most efficient. 
 
\begin{table}[h!]
  \caption{Comparison of the vaccination strategy rankings.}
  \label{tab:kendalltau}
  \begin{tabular}{|c|c|c|c|c|c|c|}
    \hline
                & \multicolumn{2}{c|}{InVS13} & \multicolumn{2}{c|}{InVS15} & \multicolumn{2}{c|}{LH10} \\
    \hline
                & $P$   & $n$   & $P$   & $n$   & $P$   & $n$   \\
    \hline
    Sampling 1  & 0.515 & 0.235 & 0.377 & 0.766 & 0.397 & 0.412 \\
    Sampling 2  & 0.324 & 0.382 & 0.377 & 0.481 & 0.456 & 0.412 \\
    Sampling 3  & 0.279 & 0.235 & 0.394 & 0.706 & 0.324 & 0.426 \\
    \hline
    \hline
                & \multicolumn{2}{c|}{LyonSchool} & \multicolumn{2}{c|}{SFHH} & \multicolumn{2}{c|}{Thiers13} \\
    \hline
                & $P$    & $n$    & $P$   & $n$   & $P$    & $n$    \\
    \hline
    Sampling 1  & -0.012 & -0.051 & 0.485 & 0.818 &  0.048 &  0.299 \\
    Sampling 2  &  0.091 &  0.083 & 0.758 & 0.485 & -0.074 & -0.108 \\
    Sampling 3  &  0.020 &  0.162 & 0.636 & 0.545 &  0.299 & -0.108 \\
    \hline
  \end{tabular}

  For each sampling method we compute Kendall's tau between the list of vaccination strategies ranked by increasing efficiency for the contact data and for the sampled co-presence networks, both in terms of the fraction of large outbreaks ($P$) and of median sizes of the large outbreaks ($n$).
\end{table}

Results are once again uneven and context dependent (see also Table \ref{tab:kendalltau}). In several cases such as SFHH the ranking of strategies obtained from the sampled co-presence is overall respected (Kendall's tau of 0.818 for the sampling method 1 on the size of outbreaks), while it can be strongly reshuffled in other cases (for instance in the Thiers13 case).

\section{Discussion and conclusion}

In this paper, we have investigated whether low resolution co-presence information can be used as a substitute for detailed face-to-face proximity data, both from the point of view of extracting large-scale structural and statistical features of the temporal contact network in a population and in data-driven models of epidemic processes in a population. We have considered several data sets collected in various contexts that contain both high-resolution data on face-to-face contacts between individuals and a coarser location data, both with temporal resolution. The location data can thus be transformed into a co-presence temporal network between individuals. Given its lower spatial resolution, this co-presence data contains much more events than the contact data, leading to much denser aggregated networks: indeed, all individuals in a given area are considered as co-present, while only some of them are typically engaged in a face-to-face contact. Despite this expected issue, a number of properties related to group structure and statistical distributions of temporal properties are similar in contact and co-presence data, with similar matrices of densities of links between groups and broad distributions of   
(aggregate) contact durations.

We have thus examined several methods to downsample the co-presence networks to create surrogate contact networks with overall the same amount of contact time than the real contact data. The surrogate data statistics are in general closer to the real contact data than the raw co-presence, in particular regarding the distribution of node degrees and link weights (and their evolution in networks aggregated over increasing time windows). These results mean in particular that the distribution of aggregate contact durations, a very important property that has a strong impact on the unfolding of processes on networks such as epidemic processes, could be approximately retrieved from simple sampling processes of the co-presence data and thus fed into data-driven models of populations. Several other properties, such as precise value of the average degree, average clustering or size of largest cliques and cores, turn out however to be strongly context-dependent. Moreover, the most central nodes of the contact network are not better identified than using the bare co-presence information.

We have moreover investigated the use of such surrogate contact data in numerical simulations of spreading processes in a population. Overall, simulations performed on surrogate data obtained with one of the sampling method yield results close to the ones obtained with the real data, while the other methods over- or under-estimate these results, but the best method turns out to depend on context (Note however that all these methods give obviously results much closer to the one of the real contact network than if raw co-presence is used, given co-presence overestimates strongly the contacts and thus yields a strongly overestimated epidemic risk). We moreover investigated the possibility to rank containment strategies according to their efficiency, and found that this ranking is once again context dependent: in some cases, simulations on sampled co-presence networks allow us to uncover the most efficient vaccination strategies for containing a spread on the real contact data, while in other cases the rankings differ quite strongly.

In conclusion, we showed that co-presence data, while yielding interesting insights into some of the large scale properties of the contact network, is not easily usable to build in a reliable and systematic fashion surrogate contact data that reproduces detailed features of the real contacts and could be used in numerical simulations to predict the outcome of spreading processes and the impact of containment strategies, at least for processes involving contagion at short distances \cite{Stopczynski:2015} (note that, while more sophisticated sampling procedures might be devised, they would most probably involve more parameters and/or more additional information not present in the raw co-presence data, and would also most probably still give context-dependent results). We note however that even coarse location information has been shown to be a useful additional information whenever the precise contact data is incomplete \cite{Sapienza:2017}. Optimally, data collection with wearable sensors should thus contain both high resolution data about relative positions of individuals, in order to detect face-to-face proximity, and coarser co-presence information to inform for instance on mobility patterns within buildings or complement potential data losses.

%%%%%%%%%%%%%%%%%%%%%%%%%%%%%%%%%%%%%%%%%%%%%%
%%                                          %%
%% Backmatter begins here                   %%
%%                                          %%
%%%%%%%%%%%%%%%%%%%%%%%%%%%%%%%%%%%%%%%%%%%%%%

\begin{backmatter}

\section*{Availability of data and material}
The datasets supporting the conclusions of this article are available on the SocioPatterns website: \url{http://www.sociopatterns.org/datasets/} (see also references in this paper).

\section*{Competing interests}
The authors declare that they have no competing interests.

%\section*{Funding}
%\red{Add funding information if necessary.}

\section*{Author's contributions}
MG and AB conceived and designed the study. MG performed the numerical simulations and statistical analysis, created the figures and wrote the first draft of the manuscript. MG and AB wrote the final version of the manuscript.

% \section*{Acknowledgements}
% Text for this section \ldots

%%  After submission of the .TEX file,                     %%
%%  you will be prompted to submit your .BBL file.         %%

% if your bibliography is in bibtex format, use those commands:

\newcommand{\BMCxmlcomment}[1]{}

\BMCxmlcomment{

<refgrp>

<bibl id="B1">
  <title><p>Social Contacts and Mixing Patterns Relevant to the Spread of
  Infectious Diseases</p></title>
  <aug>
    <au><snm>Mossong</snm><fnm>J</fnm></au>
    <au><snm>Hens</snm><fnm>N</fnm></au>
    <au><snm>Jit</snm><fnm>M</fnm></au>
    <au><snm>Beutels</snm><fnm>P</fnm></au>
    <au><snm>Auranen</snm><fnm>K</fnm></au>
    <au><snm>Mikolajczyk</snm><fnm>R</fnm></au>
    <au><snm>Massari</snm><fnm>M</fnm></au>
    <au><snm>Salmaso</snm><fnm>S</fnm></au>
    <au><snm>Tomba</snm><fnm>GS</fnm></au>
    <au><snm>Wallinga</snm><fnm>J</fnm></au>
    <au><snm>Heijne</snm><fnm>J</fnm></au>
    <au><snm>Sadkowska Todys</snm><fnm>M</fnm></au>
    <au><snm>Rosinska</snm><fnm>M</fnm></au>
    <au><snm>Edmunds</snm><fnm>WJ</fnm></au>
  </aug>
  <source>PLOS Medicine</source>
  <publisher>Public Library of Science</publisher>
  <pubdate>2008</pubdate>
  <volume>5</volume>
  <issue>3</issue>
  <fpage>1</fpage>
  <lpage>1</lpage>
  <url>https://doi.org/10.1371/journal.pmed.0050074</url>
</bibl>

<bibl id="B2">
  <title><p>Social encounter networks: characterizing Great Britain</p></title>
  <aug>
    <au><snm>Danon</snm><fnm>L</fnm></au>
    <au><snm>Read</snm><fnm>JM</fnm></au>
    <au><snm>House</snm><fnm>TA</fnm></au>
    <au><snm>Vernon</snm><fnm>MC</fnm></au>
    <au><snm>Keeling</snm><fnm>MJ</fnm></au>
  </aug>
  <source>Proc. R. Soc. B</source>
  <pubdate>2013</pubdate>
  <volume>280</volume>
  <fpage>20131037</fpage>
</bibl>

<bibl id="B3">
  <title><p>Six challenges in measuring contact networks for use in
  modelling</p></title>
  <aug>
    <au><snm>Eames</snm><fnm>K.</fnm></au>
    <au><snm>Bansal</snm><fnm>S.</fnm></au>
    <au><snm>Frost</snm><fnm>S.</fnm></au>
    <au><snm>Riley</snm><fnm>S.</fnm></au>
  </aug>
  <source>Epidemics</source>
  <pubdate>2015</pubdate>
  <volume>10</volume>
  <fpage>72</fpage>
  <lpage>77</lpage>
</bibl>

<bibl id="B4">
  <title><p>Reality mining: sensing complex social systems</p></title>
  <aug>
    <au><snm>Eagle</snm><fnm>N</fnm></au>
    <au><snm>(Sandy) Pentland</snm><fnm>A</fnm></au>
  </aug>
  <source>Personal and Ubiquitous Computing</source>
  <pubdate>2006</pubdate>
  <volume>10</volume>
  <issue>4</issue>
  <fpage>255</fpage>
  <lpage>-268</lpage>
  <url>https://doi.org/10.1007/s00779-005-0046-3</url>
</bibl>

<bibl id="B5">
  <title><p>Instrumenting the city: Developing methods for observing and
  understanding the digital cityscape</p></title>
  <aug>
    <au><snm>O’Neill</snm><fnm>E</fnm></au>
    <au><snm>Kostakos</snm><fnm>V</fnm></au>
    <au><snm>Kindberg</snm><fnm>T</fnm></au>
    <au><snm>Schiek</snm><fnm>A</fnm></au>
    <au><snm>Penn</snm><fnm>A</fnm></au>
    <au><snm>Fraser</snm><fnm>D</fnm></au>
    <au><snm>Jones</snm><fnm>T</fnm></au>
  </aug>
  <source>UbiComp 2006: Ubiquitous Computing</source>
  <publisher>Springer</publisher>
  <pubdate>2006</pubdate>
  <fpage>315</fpage>
  <lpage>-332</lpage>
</bibl>

<bibl id="B6">
  <title><p>Description and simulation of dynamic mobility networks</p></title>
  <aug>
    <au><snm>Scherrer</snm><fnm>A</fnm></au>
    <au><snm>Borgnat</snm><fnm>P</fnm></au>
    <au><snm>Fleury</snm><fnm>E</fnm></au>
    <au><snm>Guillaume</snm><fnm>J L</fnm></au>
    <au><snm>Robardet</snm><fnm>C</fnm></au>
  </aug>
  <source>Computer Networks</source>
  <publisher>Elsevier</publisher>
  <pubdate>2008</pubdate>
  <volume>52</volume>
  <issue>15</issue>
  <fpage>2842</fpage>
  <lpage>-2858</lpage>
</bibl>

<bibl id="B7">
  <title><p>Joint Bluetooth/Wifi Scanning Framework for Characterizing and
  Leveraging People Movement in University Campus</p></title>
  <aug>
    <au><snm>Vu</snm><fnm>L</fnm></au>
    <au><snm>Nahrstedt</snm><fnm>K</fnm></au>
    <au><snm>Retika</snm><fnm>S</fnm></au>
    <au><snm>Gupta</snm><fnm>I</fnm></au>
  </aug>
  <source>Proceedings of the 13th ACM International Conference on Modeling,
  Analysis, and Simulation of Wireless and Mobile Systems</source>
  <publisher>New York, NY, USA: ACM</publisher>
  <series><title><p>MSWIM '10</p></title></series>
  <pubdate>2010</pubdate>
  <fpage>257</fpage>
  <lpage>-265</lpage>
  <url>http://doi.acm.org/10.1145/1868521.1868563</url>
</bibl>

<bibl id="B8">
  <title><p>Towards a temporal network analysis of interactive WiFi
  users</p></title>
  <aug>
    <au><snm>Zhang</snm><fnm>Y</fnm></au>
    <au><snm>Wang</snm><fnm>L</fnm></au>
    <au><snm>Zhang</snm><fnm>YQ</fnm></au>
    <au><snm>Li</snm><fnm>X</fnm></au>
  </aug>
  <source>EPL (Europhysics Letters)</source>
  <pubdate>2012</pubdate>
  <volume>98</volume>
  <issue>6</issue>
  <fpage>68002</fpage>
  <url>http://stacks.iop.org/0295-5075/98/i=6/a=68002</url>
</bibl>

<bibl id="B9">
  <title><p>Measuring Large-Scale Social Networks with High
  Resolution</p></title>
  <aug>
    <au><snm>Stopczynski</snm><fnm>A</fnm></au>
    <au><snm>Sekara</snm><fnm>V</fnm></au>
    <au><snm>Sapiezynski</snm><fnm>P</fnm></au>
    <au><snm>Cuttone</snm><fnm>A</fnm></au>
    <au><snm>Madsen</snm><fnm>MM</fnm></au>
    <au><snm>Larsen</snm><fnm>JE</fnm></au>
    <au><snm>Lehmann</snm><fnm>S</fnm></au>
  </aug>
  <source>PLOS ONE</source>
  <publisher>Public Library of Science</publisher>
  <pubdate>2014</pubdate>
  <volume>9</volume>
  <issue>4</issue>
  <fpage>1</fpage>
  <lpage>24</lpage>
  <url>https://doi.org/10.1371/journal.pone.0095978</url>
</bibl>

<bibl id="B10">
  <title><p>Social sensors for automatic data collection</p></title>
  <aug>
    <au><snm>Olgu{\'\i}n</snm><fnm>DO</fnm></au>
    <au><snm>Pentland</snm><fnm>AS</fnm></au>
  </aug>
  <source>AMCIS 2008 Proceedings</source>
  <pubdate>2008</pubdate>
  <fpage>171</fpage>
</bibl>

<bibl id="B11">
  <title><p>Flunet: Automated tracking of contacts during flu
  season</p></title>
  <aug>
    <au><snm>Hashemian</snm><fnm>M. S.</fnm></au>
    <au><snm>Stanley</snm><fnm>K. G.</fnm></au>
    <au><snm>Osgood</snm><fnm>N.</fnm></au>
  </aug>
  <source>8th International Symposium on Modeling and Optimization in Mobile,
  Ad Hoc, and Wireless Networks</source>
  <pubdate>2010</pubdate>
  <fpage>348</fpage>
  <lpage>353</lpage>
</bibl>

<bibl id="B12">
  <title><p>A high-resolution human contact network for infectious disease
  transmission</p></title>
  <aug>
    <au><snm>Salath\'e</snm><fnm>M</fnm></au>
    <au><snm>Kazandjieva</snm><fnm>M</fnm></au>
    <au><snm>Lee</snm><fnm>JW</fnm></au>
    <au><snm>Levis</snm><fnm>P</fnm></au>
    <au><snm>Feldman</snm><fnm>MW</fnm></au>
    <au><snm>Jones</snm><fnm>JH</fnm></au>
  </aug>
  <source>Proceedings of the National Academy of Sciences</source>
  <pubdate>2010</pubdate>
  <volume>107</volume>
  <issue>51</issue>
  <fpage>22020</fpage>
  <lpage>22025</lpage>
  <url>http://www.pnas.org/content/107/51/22020.abstract</url>
</bibl>

<bibl id="B13">
  <title><p>Dynamics of Person-to-Person Interactions from Distributed {RFID}
  Sensor Networks</p></title>
  <aug>
    <au><snm>Cattuto</snm><fnm>C</fnm></au>
    <au><snm>{Van den Broeck}</snm><fnm>W</fnm></au>
    <au><snm>Barrat</snm><fnm>A</fnm></au>
    <au><snm>Colizza</snm><fnm>V</fnm></au>
    <au><snm>Pinton</snm><fnm>{Jean-François}</fnm></au>
    <au><snm>Vespignani</snm><fnm>A</fnm></au>
  </aug>
  <source>PLOS ONE</source>
  <publisher>Public Library of Science</publisher>
  <pubdate>2010</pubdate>
  <volume>5</volume>
  <issue>7</issue>
  <fpage>e11596</fpage>
  <url>http://dx.doi.org/10.1371%2Fjournal.pone.0011596</url>
</bibl>

<bibl id="B14">
  <title><p>Objective Measurement of Sociability and Activity: Mobile Sensing
  in the Community</p></title>
  <aug>
    <au><snm>Berke</snm><fnm>EM</fnm></au>
    <au><snm>Choudhury</snm><fnm>T</fnm></au>
    <au><snm>Ali</snm><fnm>S</fnm></au>
    <au><snm>Rabbi</snm><fnm>M</fnm></au>
  </aug>
  <source>The Annals of Family Medicine</source>
  <pubdate>2011</pubdate>
  <volume>9</volume>
  <issue>4</issue>
  <fpage>344</fpage>
  <lpage>350</lpage>
  <url>http://www.annfammed.org/content/9/4/344.abstract</url>
</bibl>

<bibl id="B15">
  <title><p>Electronic Sensors for Assessing Interactions between Healthcare
  Workers and Patients under Airborne Precautions</p></title>
  <aug>
    <au><snm>Lucet</snm><fnm>JC</fnm></au>
    <au><snm>Laouenan</snm><fnm>C</fnm></au>
    <au><snm>Chelius</snm><fnm>G</fnm></au>
    <au><snm>Veziris</snm><fnm>N</fnm></au>
    <au><snm>Lepelletier</snm><fnm>D</fnm></au>
    <au><snm>Friggeri</snm><fnm>A</fnm></au>
    <au><snm>Abiteboul</snm><fnm>D</fnm></au>
    <au><snm>Bouvet</snm><fnm>E</fnm></au>
    <au><snm>Mentre</snm><fnm>F</fnm></au>
    <au><snm>Fleury</snm><fnm>E</fnm></au>
  </aug>
  <source>PLOS ONE</source>
  <publisher>Public Library of Science</publisher>
  <pubdate>2012</pubdate>
  <volume>7</volume>
  <issue>5</issue>
  <fpage>1</fpage>
  <lpage>7</lpage>
  <url>https://doi.org/10.1371/journal.pone.0037893</url>
</bibl>

<bibl id="B16">
  <title><p>Using {Sensor} {Networks} to {Study} the {Effect} of {Peripatetic}
  {Healthcare} {Workers} on the {Spread} of {Hospital}-{Associated}
  {Infections}</p></title>
  <aug>
    <au><snm>Hornbeck</snm><fnm>T</fnm></au>
    <au><snm>Naylor</snm><fnm>D</fnm></au>
    <au><snm>Segre</snm><fnm>AM</fnm></au>
    <au><snm>Thomas</snm><fnm>G</fnm></au>
    <au><snm>Herman</snm><fnm>T</fnm></au>
    <au><snm>Polgreen</snm><fnm>PM</fnm></au>
  </aug>
  <source>The Journal of Infectious Diseases</source>
  <pubdate>2012</pubdate>
  <volume>206</volume>
  <issue>10</issue>
  <fpage>1549</fpage>
  <lpage>-1557</lpage>
  <url>http://www.ncbi.nlm.nih.gov/pmc/articles/PMC3475631/</url>
</bibl>

<bibl id="B17">
  <title><p>Measuring Social Contacts in the Emergency Department</p></title>
  <aug>
    <au><snm>Lowery North</snm><fnm>DW</fnm></au>
    <au><snm>Hertzberg</snm><fnm>VS</fnm></au>
    <au><snm>Elon</snm><fnm>L</fnm></au>
    <au><snm>Cotsonis</snm><fnm>G</fnm></au>
    <au><snm>Hilton</snm><fnm>SA</fnm></au>
    <au><snm>Vaughns</snm><fnm>CF</fnm></au>
    <au><snm>Hill</snm><fnm>E</fnm></au>
    <au><snm>Shrestha</snm><fnm>A</fnm></au>
    <au><snm>Jo</snm><fnm>A</fnm></au>
    <au><snm>Adams</snm><fnm>N</fnm></au>
  </aug>
  <source>PLOS ONE</source>
  <publisher>Public Library of Science</publisher>
  <pubdate>2013</pubdate>
  <volume>8</volume>
  <issue>8</issue>
  <fpage>1</fpage>
  <lpage>9</lpage>
  <url>https://doi.org/10.1371/journal.pone.0070854</url>
</bibl>

<bibl id="B18">
  <title><p>The role of heterogeneity in contact timing and duration in network
  models of influenza spread in schools</p></title>
  <aug>
    <au><snm>Toth</snm><fnm>DJA</fnm></au>
    <au><snm>Leecaster</snm><fnm>M</fnm></au>
    <au><snm>Pettey</snm><fnm>WBP</fnm></au>
    <au><snm>Gundlapalli</snm><fnm>AV</fnm></au>
    <au><snm>Gao</snm><fnm>H</fnm></au>
    <au><snm>Rainey</snm><fnm>JJ</fnm></au>
    <au><snm>Uzicanin</snm><fnm>A</fnm></au>
    <au><snm>Samore</snm><fnm>MH</fnm></au>
  </aug>
  <source>Journal of The Royal Society Interface</source>
  <publisher>The Royal Society</publisher>
  <pubdate>2015</pubdate>
  <volume>12</volume>
  <issue>108</issue>
  <fpage>20150279</fpage>
</bibl>

<bibl id="B19">
  <title><p>Social Contact Networks and Mixing among Students in {K-12} Schools
  in {Pittsburgh, PA}</p></title>
  <aug>
    <au><snm>Guclu</snm><fnm>H</fnm></au>
    <au><snm>Read</snm><fnm>J</fnm></au>
    <au><snm>Vukotich</snm><fnm>CJ</fnm></au>
    <au><snm>Galloway</snm><fnm>DD</fnm></au>
    <au><snm>Gao</snm><fnm>H</fnm></au>
    <au><snm>Rainey</snm><fnm>JJ</fnm></au>
    <au><snm>Uzicanin</snm><fnm>A</fnm></au>
    <au><snm>Zimmer</snm><fnm>SM</fnm></au>
    <au><snm>Cummings</snm><fnm>DAT</fnm></au>
  </aug>
  <source>PLOS ONE</source>
  <publisher>Public Library of Science</publisher>
  <pubdate>2016</pubdate>
  <volume>11</volume>
  <issue>3</issue>
  <fpage>1</fpage>
  <lpage>19</lpage>
  <url>https://doi.org/10.1371/journal.pone.0151139</url>
</bibl>

<bibl id="B20">
  <title><p>The importance of including dynamic social networks when modeling
  epidemics of airborne infections: does increasing complexity increase
  accuracy?</p></title>
  <aug>
    <au><snm>Blower</snm><fnm>S</fnm></au>
    <au><snm>Go</snm><fnm>MH</fnm></au>
  </aug>
  <source>BMC Medicine</source>
  <pubdate>2011</pubdate>
  <volume>9</volume>
  <issue>1</issue>
  <fpage>88</fpage>
  <url>http://www.biomedcentral.com/1741-7015/9/88</url>
</bibl>

<bibl id="B21">
  <title><p>Measuring contact patterns with wearable sensors: methods, data
  characteristics and applications to data-driven simulations of infectious
  diseases</p></title>
  <aug>
    <au><snm>Barrat</snm><fnm>A.</fnm></au>
    <au><snm>Cattuto</snm><fnm>C.</fnm></au>
    <au><snm>Tozzi</snm><fnm>A. E.</fnm></au>
    <au><snm>Vanhems</snm><fnm>P.</fnm></au>
    <au><snm>Voirin</snm><fnm>N.</fnm></au>
  </aug>
  <source>Clinical Microbiology and Infection</source>
  <pubdate>2014</pubdate>
  <volume>20</volume>
  <issue>1</issue>
  <fpage>10</fpage>
  <lpage>-16</lpage>
</bibl>

<bibl id="B22">
  <title><p>Inferring Relevant Social Networks from Interpersonal
  Communication</p></title>
  <aug>
    <au><snm>De Choudhury</snm><fnm>M</fnm></au>
    <au><snm>Mason</snm><fnm>WA</fnm></au>
    <au><snm>Hofman</snm><fnm>JM</fnm></au>
    <au><snm>Watts</snm><fnm>DJ</fnm></au>
  </aug>
  <source>Proceedings of the 19th International Conference on World Wide
  Web</source>
  <publisher>New York, NY, USA: ACM</publisher>
  <series><title><p>WWW '10</p></title></series>
  <pubdate>2010</pubdate>
  <fpage>301</fpage>
  <lpage>-310</lpage>
  <url>http://doi.acm.org/10.1145/1772690.1772722</url>
</bibl>

<bibl id="B23">
  <title><p>Inferring friendship network structure by using mobile phone
  data</p></title>
  <aug>
    <au><snm>Eagle</snm><fnm>N</fnm></au>
    <au><snm>Pentland</snm><fnm>AS</fnm></au>
    <au><snm>Lazer</snm><fnm>D</fnm></au>
  </aug>
  <source>Proceedings of the National Academy of Sciences</source>
  <pubdate>2009</pubdate>
  <volume>106</volume>
  <issue>36</issue>
  <fpage>15274</fpage>
  <lpage>15278</lpage>
  <url>http://www.pnas.org/content/106/36/15274.abstract</url>
</bibl>

<bibl id="B24">
  <title><p>Inferring social ties from geographic coincidences</p></title>
  <aug>
    <au><snm>Crandall</snm><fnm>DJ</fnm></au>
    <au><snm>Backstrom</snm><fnm>L</fnm></au>
    <au><snm>Cosley</snm><fnm>D</fnm></au>
    <au><snm>Suri</snm><fnm>S</fnm></au>
    <au><snm>Huttenlocher</snm><fnm>D</fnm></au>
    <au><snm>Kleinberg</snm><fnm>J</fnm></au>
  </aug>
  <source>Proceedings of the National Academy of Sciences</source>
  <pubdate>2010</pubdate>
  <volume>107</volume>
  <issue>52</issue>
  <fpage>22436</fpage>
  <lpage>22441</lpage>
  <url>http://www.pnas.org/content/107/52/22436.abstract</url>
</bibl>

<bibl id="B25">
  <title><p>New insights and methods for predicting face-to-face
  contacts</p></title>
  <aug>
    <au><snm>Scholz</snm><fnm>C</fnm></au>
    <au><snm>Atzmueller</snm><fnm>M</fnm></au>
    <au><snm>Stumme</snm><fnm>G</fnm></au>
    <au><snm>Barrat</snm><fnm>A</fnm></au>
    <au><snm>Cattuto</snm><fnm>C</fnm></au>
  </aug>
  <source>7th International Conference on Weblogs and Social Media
  (ICWSM-13)</source>
  <pubdate>2013</pubdate>
</bibl>

<bibl id="B26">
  <title><p>Inferring Person-to-person Proximity Using WiFi Signals</p></title>
  <aug>
    <au><snm>Sapiezynski</snm><fnm>P</fnm></au>
    <au><snm>Stopczynski</snm><fnm>A</fnm></au>
    <au><snm>Wind</snm><fnm>DK</fnm></au>
    <au><snm>Leskovec</snm><fnm>J</fnm></au>
    <au><snm>Lehmann</snm><fnm>S</fnm></au>
  </aug>
  <source>Proc. ACM Interact. Mob. Wearable Ubiquitous Technol.</source>
  <publisher>New York, NY, USA: ACM</publisher>
  <pubdate>2017</pubdate>
  <volume>1</volume>
  <issue>2</issue>
  <fpage>24:1</fpage>
  <lpage>-24:20</lpage>
  <url>http://doi.acm.org/10.1145/3090089</url>
</bibl>

<bibl id="B27">
  <title><p>SocioPatterns Collaboration</p></title>
  <source>\url{www.sociopatterns.org}</source>
  <note>Accessed 18 Oct 2017</note>
</bibl>

<bibl id="B28">
  <title><p>Compensating for population sampling in simulations of epidemic
  spread on temporal contact networks</p></title>
  <aug>
    <au><snm>G\'enois</snm><fnm>M</fnm></au>
    <au><snm>Vestergaard</snm><fnm>CL</fnm></au>
    <au><snm>Cattuto</snm><fnm>C</fnm></au>
    <au><snm>Barrat</snm><fnm>A</fnm></au>
  </aug>
  <source>Nature Communications</source>
  <publisher>Nature Publishing Group</publisher>
  <pubdate>2015</pubdate>
  <volume>6</volume>
  <fpage>8860</fpage>
  <url>http://dx.doi.org/10.1038/ncomms9860</url>
</bibl>

<bibl id="B29">
  <title><p>Estimating the outcome of spreading processes on networks with
  incomplete information: a mesoscale approach</p></title>
  <aug>
    <au><snm>Sapienza</snm><fnm>A</fnm></au>
    <au><snm>Barrat</snm><fnm>A</fnm></au>
    <au><snm>Cattuto</snm><fnm>C</fnm></au>
    <au><snm>Gauvin</snm><fnm>L</fnm></au>
  </aug>
  <source>ArXiv e-prints</source>
  <pubdate>2017</pubdate>
</bibl>

<bibl id="B30">
  <title><p>A low-cost method to assess the epidemiological importance of
  individuals in controlling infectious disease outbreaks</p></title>
  <aug>
    <au><snm>Smieszek</snm><fnm>T</fnm></au>
    <au><snm>Salath\'e</snm><fnm>M</fnm></au>
  </aug>
  <source>BMC MEDICINE</source>
  <pubdate>2013</pubdate>
  <volume>11</volume>
  <issue>1</issue>
  <fpage>35</fpage>
  <url>http://www.biomedcentral.com/1741-7015/11/35</url>
</bibl>

<bibl id="B31">
  <title><p>Data on face-to-face contacts in an office building suggest a
  low-cost vaccination strategy based on community linkers</p></title>
  <aug>
    <au><snm>G\'enois</snm><fnm>M</fnm></au>
    <au><snm>Vestergaard</snm><fnm>CL</fnm></au>
    <au><snm>Fournet</snm><fnm>J</fnm></au>
    <au><snm>Panisson</snm><fnm>A</fnm></au>
    <au><snm>Bonmarin</snm><fnm>I</fnm></au>
    <au><snm>Barrat</snm><fnm>A</fnm></au>
  </aug>
  <source>Network Science</source>
  <pubdate>2015</pubdate>
  <volume>3</volume>
  <fpage>326</fpage>
  <lpage>-347</lpage>
  <url>http://journals.cambridge.org/article_S2050124215000107</url>
</bibl>

<bibl id="B32">
  <title><p>Estimating Potential Infection Transmission Routes in Hospital
  Wards Using Wearable Proximity Sensors</p></title>
  <aug>
    <au><snm>Vanhems</snm><fnm>P</fnm></au>
    <au><snm>Barrat</snm><fnm>A</fnm></au>
    <au><snm>Cattuto</snm><fnm>C</fnm></au>
    <au><snm>Pinton</snm><fnm>JF</fnm></au>
    <au><snm>Khanafer</snm><fnm>N</fnm></au>
    <au><snm>R\'egis</snm><fnm>C</fnm></au>
    <au><snm>Kim</snm><fnm>Ba</fnm></au>
    <au><snm>Comte</snm><fnm>B</fnm></au>
    <au><snm>Voirin</snm><fnm>N</fnm></au>
  </aug>
  <source>PLoS ONE</source>
  <publisher>Public Library of Science</publisher>
  <pubdate>2013</pubdate>
  <volume>8</volume>
  <issue>9</issue>
  <fpage>e73970</fpage>
  <url>http://dx.doi.org/10.1371%2Fjournal.pone.0073970</url>
</bibl>

<bibl id="B33">
  <title><p>High-Resolution Measurements of Face-to-Face Contact Patterns in a
  Primary School</p></title>
  <aug>
    <au><snm>Stehl\'e</snm><fnm>J</fnm></au>
    <au><snm>Voirin</snm><fnm>N</fnm></au>
    <au><snm>Barrat</snm><fnm>A</fnm></au>
    <au><snm>Cattuto</snm><fnm>C</fnm></au>
    <au><snm>Isella</snm><fnm>L</fnm></au>
    <au><snm>Pinton</snm><fnm>JF</fnm></au>
    <au><snm>Quaggiotto</snm><fnm>M</fnm></au>
    <au><snm>Broeck</snm><fnm>W</fnm></au>
    <au><snm>R\'egis</snm><fnm>C</fnm></au>
    <au><snm>Lina</snm><fnm>B</fnm></au>
    <au><snm>Vanhems</snm><fnm>P</fnm></au>
  </aug>
  <source>PLoS ONE</source>
  <publisher>Public Library of Science</publisher>
  <pubdate>2011</pubdate>
  <volume>6</volume>
  <issue>8</issue>
  <fpage>e23176</fpage>
  <url>http://dx.doi.org/10.1371/journal.pone.0023176</url>
</bibl>

<bibl id="B34">
  <title><p>What's in a Crowd? Analysis of Face-to-Face Behavioral
  Networks</p></title>
  <aug>
    <au><snm>Isella</snm><fnm>L</fnm></au>
    <au><snm>Stehl\'e</snm><fnm>J</fnm></au>
    <au><snm>Barrat</snm><fnm>A</fnm></au>
    <au><snm>Cattuto</snm><fnm>C</fnm></au>
    <au><snm>Pinton</snm><fnm>JF</fnm></au>
    <au><snm>Broeck</snm><fnm>W</fnm></au>
  </aug>
  <source>Journal of Theoretical Biology</source>
  <pubdate>2011</pubdate>
  <volume>271</volume>
  <issue>1</issue>
  <fpage>166</fpage>
  <lpage>-180</lpage>
  <url>http://www.sciencedirect.com/science/article/B6WMD-51M60KS-2/2/cb31bee32b340b3044c724b88779a60e</url>
</bibl>

<bibl id="B35">
  <title><p>Contact Patterns in a High School: A Comparison between Data
  Collected Using Wearable Sensors, Contact Diaries and Friendship
  Surveys</p></title>
  <aug>
    <au><snm>Mastrandrea</snm><fnm>R</fnm></au>
    <au><snm>Fournet</snm><fnm>J</fnm></au>
    <au><snm>Barrat</snm><fnm>A</fnm></au>
  </aug>
  <source>PLoS ONE</source>
  <publisher>Public Library of Science</publisher>
  <pubdate>2015</pubdate>
  <volume>10</volume>
  <issue>9</issue>
  <fpage>1</fpage>
  <lpage>26</lpage>
  <url>http://dx.doi.org/10.1371%2Fjournal.pone.0136497</url>
</bibl>

<bibl id="B36">
  <title><p>Simulation of an SEIR infectious disease model on the dynamic
  contact network of conference attendees</p></title>
  <aug>
    <au><snm>Stehl{\'e}</snm><fnm>J</fnm></au>
    <au><snm>Voirin</snm><fnm>N</fnm></au>
    <au><snm>Barrat</snm><fnm>A</fnm></au>
    <au><snm>Cattuto</snm><fnm>C</fnm></au>
    <au><snm>Colizza</snm><fnm>V</fnm></au>
    <au><snm>Isella</snm><fnm>L</fnm></au>
    <au><snm>R{\'e}gis</snm><fnm>C</fnm></au>
    <au><snm>Pinton</snm><fnm>JF</fnm></au>
    <au><snm>Khanafer</snm><fnm>N</fnm></au>
    <au><snm>Broeck</snm><fnm>W</fnm></au>
    <au><snm>Vanhems</snm><fnm>P</fnm></au>
  </aug>
  <source>BMC Medicine</source>
  <pubdate>2011</pubdate>
  <volume>9</volume>
  <issue>1</issue>
  <fpage>87</fpage>
  <url>https://doi.org/10.1186/1741-7015-9-87</url>
</bibl>

<bibl id="B37">
  <title><p>The scaling of human interactions with city size</p></title>
  <aug>
    <au><snm>M.</snm><fnm>S</fnm></au>
    <au><snm>M.</snm><fnm>BL</fnm></au>
    <au><snm>S.</snm><fnm>G</fnm></au>
    <au><snm>M.</snm><fnm>R</fnm></au>
    <au><snm>R.</snm><fnm>C</fnm></au>
    <au><snm>Z.</snm><fnm>S</fnm></au>
    <au><snm>B.</snm><fnm>WG</fnm></au>
    <au><snm>C.</snm><fnm>R</fnm></au>
  </aug>
  <source>J. R. Soc. Interface</source>
  <pubdate>2014</pubdate>
  <volume>11</volume>
  <fpage>20130789</fpage>
</bibl>

<bibl id="B38">
  <title><p>The Scaling of Human Contacts and Epidemic Processes in
  Metapopulation Networks</p></title>
  <aug>
    <au><snm>Tizzoni</snm><fnm>M.</fnm></au>
    <au><snm>Sun</snm><fnm>K.</fnm></au>
    <au><snm>Benusiglio</snm><fnm>D.</fnm></au>
    <au><snm>Karsai</snm><fnm>M.</fnm></au>
    <au><snm>Perra</snm><fnm>N.</fnm></au>
  </aug>
  <source>Sci. Rep.</source>
  <pubdate>2015</pubdate>
  <volume>5</volume>
  <fpage>15111</fpage>
</bibl>

<bibl id="B39">
  <title><p>Physical Proximity and Spreading in Dynamic Social
  Networks</p></title>
  <aug>
    <au><snm>Stopczynski</snm><fnm>A.</fnm></au>
    <au><snm>Pentland</snm><fnm>A</fnm></au>
    <au><snm>Lehmann</snm><fnm>S</fnm></au>
  </aug>
  <source>preprint</source>
  <pubdate>2015</pubdate>
  <fpage>arXiv:1509.06530</fpage>
</bibl>

</refgrp>
} % end of \BMCxmlcomment
% \bibliographystyle{bmc-mathphys} % Style BST file (bmc-mathphys, vancouver, spbasic).
% \bibliography{biblio}            % Bibliography file (usually '*.bib' )
% for author-year bibliography (bmc-mathphys or spbasic)
% a) write to bib file (bmc-mathphys only)
% @settings{label, options="nameyear"}

% or include bibliography directly:
% \begin{thebibliography}
% \bibitem{b1}
% \end{thebibliography}

%%%%%%%%%%%%%%%%%%%%%%%%%%%%%%%%%%%
%%                               %%
%% Figures                       %%
%%                               %%
%% NB: this is for captions and  %%
%% Titles. All graphics must be  %%
%% submitted separately and NOT  %%
%% included in the Tex document  %%
%%                               %%
%%%%%%%%%%%%%%%%%%%%%%%%%%%%%%%%%%%

%%
%% Do not use \listoffigures as most will included as separate files

% \section*{Figures}

%%%%%%%%%%%%%%%%%%%%%%%%%%%%%%%%%%%
%%                               %%
%% Tables                        %%
%%                               %%
%%%%%%%%%%%%%%%%%%%%%%%%%%%%%%%%%%%

%% Use of \listoftables is discouraged.
%%
% \section*{Tables}

%%%%%%%%%%%%%%%%%%%%%%%%%%%%%%%%%%%
%%                               %%
%% Additional Files              %%
%%                               %%
%%%%%%%%%%%%%%%%%%%%%%%%%%%%%%%%%%%

% \section*{Additional Files}
% \subsection*{Supplementary Information --- SI.pdf}
% This file contains additional tables and figures, in particular for the other data sets considered in this paper.
% \subsection*{Data sets --- datasets.tar.bz2}
% This file contains the data used for the present paper.
\end{backmatter}

\end{document}